\documentclass[12pt,preprint]{aastex}

\slugcomment{Submitted to the Astrophysical Journal August 24, 2006, Accepted By the ApJ January 23, 2007}

\shorttitle{Young, Wide, VLM and Brown Dwarf Binaries}
\shortauthors{Close et al.}

\begin{document}


\title{The Wide Brown Dwarf Binary Oph 1622-2405 and  
Discovery of A Wide, Low Mass Binary in 
Ophiuchus (Oph 1623-2402): A New Class of Young Evaporating Wide 
Binaries?$^{0}$ 
\footnotetext{ Based on observations made with the Keck and Gemini North
telescopes}}



\author{Laird M. Close\altaffilmark{1}, B. Zuckerman\altaffilmark{2}, Inseok 
Song\altaffilmark{3}, Travis Barman\altaffilmark{4}, Christian
Marois\altaffilmark{5}, Emily L. Rice\altaffilmark{2}, 
Nick Siegler\altaffilmark{1}, Bruce Macintosh\altaffilmark{5}, E. E.
Becklin\altaffilmark{2},
Randy Campbell\altaffilmark{6}, James E.
Lyke\altaffilmark{6}, Al Conrad\altaffilmark{6}, \& David Le Mignant\altaffilmark{6}}

\email{lclose@as.arizona.edu}

\altaffiltext{1}{Steward Observatory, University of Arizona, Tucson, AZ 85721, USA}
\altaffiltext{2}{Dept. of Physics \& Astronomy and NASA Astrobiology Institute,
University of
California, Los Angeles, CA 90095, USA} 
\altaffiltext{3}{Gemini Observatory, 670 North A'ohoku Place, Hilo, HI 96720, USA}
\altaffiltext{4}{Lowell Observatory, 1400 West Mars Hill Rd., Flagstaff AZ 86001, USA}     
\altaffiltext{5}{Lawrence Livermore Lab, 7000 E. Ave, Livermore, CA 94550, USA}   
\altaffiltext{6}{W.M. Keck Observatory, Kamuela, HI 96743, USA}   



\begin{abstract} 

We imaged five objects near the star forming clouds of Ophiuchus with
the Keck Laser Guide Star AO system.  We resolved Allers et
al. (2006)'s \#11 (Oph 16222-2405) and \#16 (Oph 16233-2402) into binary 
systems. The \#11 object is resolved
into a $243$ AU binary, the widest known for a very low mass (VLM) binary. 
The binary
nature of \#11 was discovered first by Allers (2005) and independently here
during which we 
obtained the first spatially resolved $R\sim2000$ near-infrared (J \&
K) spectra, mid-IR photometry, and orbital motion estimates. We
estimate for 11A and 11B gravities (log(g)$>3.75$), ages ($5\pm2$
Myr), luminosities (log(L/$L_{\sun}$)=$-2.77\pm0.10$ and
$-2.96\pm0.10$), and temperatures ($T_{eff}=2375\pm175$ and
$2175\pm175$ K). We find self-consistent DUSTY evolutionary model
(Chabrier et al. 2000) masses of $17^{+4}_{-5} M_{Jup}$ and
$14^{+6}_{-5} M_{Jup}$, for 11A and 11B respectively. Our masses are
higher than those previously reported (13--15 $M_{Jup}$ and 7--8
$M_{Jup}$) by Jayawardhana \& Ivanov (2006b). {\it Hence, we
find the system is unlikely a ``planetary mass binary'', (in agreement with Luhman et al. 2007) but it has
the second lowest mass and lowest binding energy of any known
binary}. Oph \#11 and Oph \#16 belong to a newly recognized
population of wide ($\ga100$ AU), young ($<10$ Myr), roughly equal
mass, VLM stellar and brown dwarf binaries. We deduce that $\sim
6\pm3\%$ of young ($<10$ Myr) VLM objects are in such wide
systems. However, only $0.3\pm0.1\%$ of old field VLM objects are
found in such wide systems.  Thus, young, wide, VLM binary
populations may be evaporating, due to stellar encounters in their
natal clusters, leading to a field population depleted in wide VLM
systems.


\end{abstract}


\section {Introduction}

The system 2MASS1207334-393254 (sep$\sim$41 AU,
2M1207A$\sim$21$M_{Jup}$, 2M1207b$\sim$5$M_{Jup}$) demonstrated that very 
low-mass objects could
remain bound for at least 8 Myr (Chauvin et al.  2004; 2005c; Mamajek 
2005; Song et
al.  2006).  A somewhat similar very wide ($\sim$241 AU), very
young, nearly equal mass, brown dwarf binary (2MASS J1101192-773238)
has been discovered in Chameleon (Luhman 2004). While Billeres et
al. (2005) discovered a wide ($\sim200$ AU), much older, field
binary Denis J0551-44.  They noted that "wide ultracool binaries are
undoubtedly rare in the field". Indeed, of the $\sim$69 field VLM
binaries now known only Denis J0551-44 and Denis 2200-30 ($\sim38$ AU)
have separations wider than $\sim30$AU (Close et al. 2003; Bouy et
al. 2003; see also Burgasser et al. 2006; and Siegler et al. 2005 and
references therein).

Calculations of opacity-limited fragmentation in a turbulent 3-D
medium yield minimum masses $\sim$7 Mjup (e.g., Low \& Lynden-Bell
1976; Boyd \& Whitworth 2005, Bate 2005 and references therein).
Thus, a binary system with individual masses in the vicinity of 7
times that of Jupiter is theoretically plausible; however, it should
also be very rare since even a $>5$\% fraction of {\it higher mass}
binary brown dwarfs is difficult to produce from star formation
simulations (Bate et al. 2002; 2003).  Jayawardhana \& Ivanov
(2006b) recently claimed discovery of just such a wide
``planetary mass binary''. We independently discovered this very
same pair "Oph 11" in the course of our high-resolution survey of the Allers 
et
al. (2006) Ophiuchus dark cloud low-mass objects (we note in passing that 
Allers (2005) was the first to resolve this system).

Allers et al. (2006) present a near-infrared (1.2, 1.6, and 2.2
$\mu$m; hereafter NIR) imaging survey of several young star formation
associations.  By cross-correlation with the ``Cores to Disks (c2d)''
${\it Spitzer}$ [3.6], [4.5], [5.8], [8.0], \& [24] $\mu$m legacy
mid-IR survey (Evans et al. 2003) of these same star formation
regions, they were able to identify 19 candidate very young, low-mass,
objects each of which appears to have a significant mid-IR excess in the 
${\it Spitzer}$ dataset.

In a high spatial resolution imaging survey of the Allers et
al. (2006) Ophiuchus cloud sources, we found their sources \#11 \&
\#16 to be split into $1.7-1.9\arcsec$ binaries. In sections 4.1-4.3 
we prove that \#11AB is a physical binary. We consider the 
binary nature of \#16AB in Section 4.5.

While Allers et al. (2006) estimate a mass of $\sim9$ Jupiters
for Oph 11B, based on their 
optical spectra and NIR photometry, Jayawardhana \&
Ivanov (2006b) estimate masses of 13-15 Jupiters and 7-8 Jupiters for
11A and 11B, respectively, with the DUSTY models of Chabrier et al. (2000). We fully summarize all the recent results on Oph \#11 in section 4.4.

We will show, after an analysis of each component's luminosity, surface
gravity, age, and $T_{eff}$, that the masses of 11A and 11B are likely just
above the deuterium-burning limit ($\sim13 M_{Jup}$). While we find it
unlikely that Oph \#11 is a planetary-mass binary, we do confirm that it is
unusually wide, and of extremely low-mass, with likely the lowest binding
energy of any known brown dwarf binary.
Oph \#16 has the $4^{th}$ lowest binding energy of any known VLM system.

In section 4.6 we note that \#11 and \#16 join four other
recently discovered young, wide, VLM binaries which, we argue, define a
new class of ``young, wide, VLM binaries''. We find such wide VLM
systems to be very rare in the field but $\sim$15 times more common in very
young clusters and associations. In section 4.6.3 we suggest that
stellar encounters in these clusters have the potential to
break-up/evaporate these binaries.

\section{OBSERVATIONS \& REDUCTIONS}

\subsection{Imaging Oph \#11 and Oph \#16 at J, H, Ks, \& Ls}

We independently discovered the binary nature of the Oph 11 system
with the Gemini NIRI camera (Hodapp
et al. 2003) on July 1, 2006 (UT) in the J, H, \& Ks bands (all other objects in our survey were observed with the Keck II LGS AO system).  We then obtained 
Ls ($\sim3.0\mu$m) images with the NIRC camera on the Keck I telescope on 
July
7, 2006 (UT). In all cases seeing was excellent, yielding images with
FWHM$\sim$0.3'' in the Ks band. We reduced these data in the standard
manner (Close et al. 2002, 2003). All images were fully flat-fielded,
sky and dark subtracted, bad pixel masked, aligned, and medianed.

Figures 1 and 2 and Tables 1-3 present the measured and derived characteristics
of the Oph 11 and 16 binaries.  Since the northern component is brighter (in
the NIR) we name it Oph 11A and the apparently cooler southern component Oph
11B.

The system is also known as 2MASS J16222522-2405138 and Jayawardhana \&
Ivanov (2006b) refer to it as Oph 162225-240515 (and Oph 1622). We will refer
to this system as Oph \#11 (and as Oph 1622-2405) since it was first
mentioned in the widely available literature as \#11 by Allers et al. (2006)
to be an interesting low-mass object (in fact it was 11B that Allers was
referring to; Allers et al. 2007). 
The designation of Oph 1622 is inappropriate since Allers et al.'s
\#10, \#11, \#12, and \#13 objects would all also be ``Oph 1622'' due to
their similar RA.

The night of July 1, 2006 (UT) was photometric and zeropoints from
photometric standards were used for calibration at J, H, and Ks. There
are small ($2MASS-Gemini=$ $-0.12\pm0.09$, $0.16\pm0.10$,
$0.06\pm0.10$ mag) differences between our calibration and the J, H,
and Ks fluxes of \#11 in the 2MASS point source catalog. However, the
$1.9\arcsec$ separation of the binary likely leads to some errors in
the 2MASS $4\arcsec$ aperture photometry, and so we adopt our values
(which take the separations of the binary into account) throughout the
present paper.

We also analyzed the public ${\it Spitzer}$ [3.6], [4.5], [5.8], [8.0]
$\mu$m IRAC images of \#11. Utilizing a custom IRAC PSF fitting
program we were able to detect (for the first time) both components of
the binary in all IRAC passbands (see, for example, Figure 3). In all
cases a double PSF model of a 1.94$\arcsec$ binary with
PA=$176^{\circ}$ was a significantly better fit to the IRAC images
than a single IRAC PSF.

\subsection{NIRSPEC J and H band Spectra of 11A and 11B}

Spatially resolved $J$- and $K$-band spectra of the \#11 system
components were obtained on August 9, 2006 with NIRSPEC in low resolution 
mode on the Keck~II
telescope (Mclean et al. 1998, 2000).  We used a
0$\farcs$57$\times$42\arcsec~slit and the image rotator to maintain
the $\sim$180\degr~ position angle with both components on the slit.
The 0$\farcs$57-wide slit corresponds to 3 pixels on the detector and
a nominal resolving power of R$\sim$2000.  $J$-band observations
utilized the NIRSPEC-3 order-sorting filter (1.143$-$1.375~$\micron$),
and $K$-band observations used the $K$ filter
(1.996$-$2.382~$\micron$).  For each filter a set of four frames was
obtained in an ABBA nod pattern, nodding $\sim$20\arcsec~along the
slit between the A and B positions.  Integration times for each
exposure were 360~seconds in the $J$-band and 180~seconds in the
$K$-band.  An A0V calibrator star, HIP 80224, was observed in each
band for a 60 second integration at each nod position and at a similar
airmass to facilitate the removal of telluric absorption features from
the target spectra.  Flat, dark, and arc lamp exposures were obtained
with the calibration unit internal to NIRSPEC.

The NIRSPEC data were reduced using the REDSPEC code,\footnote{See
http://www2/keck.hawaii.edu/inst/nirspec/redspec} a package of IDL
procedures developed specifically for the reduction of NIRSPEC spectra
by S. Kim, L. Prato, and I. S. McLean.  The reduction of low resolution
NIRSPEC data with REDSPEC is described in detail in McLean et
al. (2003), and summarized here.  REDSPEC uses arc lamp frames to
spatially and spectrally rectify the raw data and to determine a
wavelength solution from observed neon and argon emission lines of
known wavelengths.  Following rectification, pairs of target nods are
subtracted to remove background and divided by the rectified,
dark-subtracted flat field.  The target and calibrator spectra are
extracted by summing over 9-11 rows of data centered on the peak of
the trace.  For the Oph~11 observations, the peaks of the traces are
separated by 18 pixels, resulting in minimum contamination between
components.  Intrinsic spectral features of the calibrator (Pa$\beta$
at 1.282~$\micron$ in the $J$-band and Br$\gamma$ at 2.166~$\micron$
in the $K$-band) are removed by linear interpolation over the line.
The calibrator spectrum is divided from the target spectrum, and
the target spectrum is multiplied by a 9770~K blackbody (the $T_{eff}$
of HIP~80224) to restore the spectral slope.  To produce the final
spectrum, the extracted spectra from all four nods are averaged and
normalized to the continuum level at a given wavelength.

The actual resolution of the observations was estimated by fitting a
Gaussian to several emission lines in the raw arc lamp frames, which
resulted in R$\sim$1500 for $J$-band and R$\sim$1900 for $K$-band.
The signal-to-noise is calculated from the maximum peak-to-peak
variation and estimated to be S/N~$\gtrsim$~100 for each spectrum.

\section{ANALYSIS}

\subsection{Extinction}

The JHK colors from Table 1 are bluer in H-K and redder in J-H than
the stellar locus of old M \& L dwarfs (see Figure 4).  Kirkpatrick et
al (2006) discuss the peculiar shape of the H-band in the spectra of
low gravity objects.  These colors are not due to a highly (line of
sight) extincted source since the visible spectrum of \#11 was
measured to be that of a young M9 by Jayawardhana \& Ivanov (2006a)
and they (as do Allers et al. 2006; 2007) find A$_{V}$=0 is a good fit
to the system. Moreover, our own comparison of the optical spectra of
11B from Jayawardhana \& Ivanov (2006b) to a young L0 (2MASS
J01415823-4633574; Kirkpatrick et al. 2006) also suggests the
extinction towards 11A and 11B must be very low.  Consistent with a
low A$_{V}$, we note there is little 24 $\mu$m IR cirrus in the
immediate area around \#11.  This is not surprising as \#11 is located
($dist\sim0.5^{\circ}\sim1pc$) near the edge of the $^{13}$CO core of
the $\rho$ Oph cloud, although \#11 (and \#16) are still inside the
$\rho$ Oph survey area (and the $^{13}$CO cloud) of Wilking et al. (2005). 

\subsection{J and K Band Spectra of 11B}

From the mainly gravity independent, yet temperature sensitive, K-band
CO lines (see for example Gorlova et al. 2003) we can measure
$T_{eff}$ and spectral type for 11A and 11B. However, this first
requires some calibrated, young, cool brown dwarfs for
comparison. There does not yet exist a standard set of very young,
late M/early L spectral standards. However, the young
brown dwarf 2MASS J01415823-4633574 (hereafter 2M0141) has been
studied with relatively high resolution spectra from 0.5-2.5 $\mu$m by
Kirkpatrick et al. (2006). They find that 2M0141 is best fit by an L0
spectral type and by $T_{eff}=2000\pm100$K and log(g)=$4.0\pm0.5$ from
detailed optical to NIR spectral synthesis fits. Hence, this is an
obvious brown dwarf to compare our spectra to (see Figures 5--8). In
addition, McGovern et al. (2004) obtained some J-band spectra of a
series of young brown dwarfs with a similar instrumental
configuration (hence spectral resolution) of NIRSPEC as we did. Hence,
we will also compare our J-band spectra to theirs.

\subsubsection{The Temperature of 11B}

In Figure 6 we compare 11B to 2M0141. The excellent
match of 11B's CO, NaI, CaII, and pseudo-continuum to the K-band
spectrum of 2M0141 suggests an $T_{eff}\sim2000$K and $\sim5$Myr
age (log(g)=4).

When we compare 11B in the J-band to 2M0141 (middle trace of Figure
8), 11B appears slightly too hot for a good fit. In fact, a better fit
to the J-band of 11B is the young ($5\pm2$ Myr) somewhat hotter (M9)
brown dwarf $\sigma$ Ori 51 (McGovern et al. 2004). Hence, it appears
that in the J-band 11B appears closer to that of an M9. Uncertainty at
the level of one subclass for young brown dwarfs is not unusual. Since
the $T_{eff}$ for an M9 is 2300K from the temperature scale of Martin
et al. (1999), 2400K from that of Luhman et al. (1999), and 2400K from
Golomoski et al. (2004) we adopt a value of 2350K for M9. So it
appears 11B could be as hot as $T_{eff}=2350$K from its J-band
spectrum (but as cool as 2000K in the K-band). Therefore, we adopt
$T_{eff}=2175\pm175$ (M$9.5\pm0.5$) as a reasonable match to the J \&
K band spectra of 11B compared to other young brown dwarfs. We note
this is consistent with the M9-L0 spectral type found by Jayawardhana
\& Ivanov (2006b) from the visible spectrum of 11B.

\subsubsection{The Surface Gravity and Age of 11B}

In both the J and K spectra, log(g) appears consistent with $\sim4.0$
and age $5\pm2$Myr. Younger ages seem to be precluded for 11B due
to the poor fit of 11B to the very young (1--2 Myr) KPNO Tau-4 brown
dwarf. Since the M9.5 spectral type of KPNO Tau-4 is consistent with
the M9-L0 of 11B, we reason that the very poor fit of 11B is primarily
due to significantly higher gravity in 11B compared to KPNO Tau-4. Indeed
it is well known that the J-band has a host of gravity sensitive
features (like the KI doublets and its entire pseudo-continuum;
McGovern et al. 2004), hence we can reconcile the poor fit of 11B to
KPNO Tau-4 and the good fit to $\sigma$ Ori 51 by adopting an age of
$5\pm2$ (hence surface gravities of log(g)$>3.75$ according to the
DUSTY models) for the Oph 11 system. These ages are significantly
higher than the 1 Myr assumed by Jayawardhana \& Ivanov
(2006b). 

However, older ages of $\sim30$Myr are very unlikely since
the strength of the NaI and CaI lines in 11B (Fig. 5 \& 6) are much weaker
than that of GSC 8047 (a known 30 Myr old M9.5 Tuc association member;
Chauvin et al. 2005b).

\subsubsection{The Temperature and Age of 11A}

We find 11A is consistently $\sim$200K hotter than 11B from
detailed spectral synthesis modeling of the J-band and K-band spectra
(section 3.3.1). Moreover, M8.5-M9.5 spectra of age $5\pm2$ Myr should
be a reasonable fit to 11A's slightly hotter spectra than the
standards of Figs 5 \& 7. Hence, a $T_{eff}\sim2375\pm175$K (or
M$9\pm0.5$) is adopted for 11A. Again our NIR spectral type is
consistent with the M9 spectral type found by Jayawardhana \& Ivanov
(2006b) from 11A's visible spectrum. Although it is somewhat hard to
disentangle the effects of temperature and surface gravity, the
similarity of 11A's spectra to that of 11B suggests that the age of 11A 
is consistent with $5\pm2$ Myr.

\subsection{Synthetic Spectral Fits}   

\subsubsection{Temperatures}

We consider how our adopted $\sim$5 Myr age and M9 and
M9.5 spectral type estimates compare to synthetic spectra with
log(g)$\sim$3.95 (consistent with the DUSTY model's 5 Myr isochrone)
and $T_{eff}$=2375K for 11A and $T_{eff}$=2175K for 11B. In Figure 9
we compare our observed K-band spectra to synthetic spectra computed
using up-to-date {\tt PHOENIX} DUSTY atmosphere models (Hauschildt et
al. in prep). The synthetic spectra do a reasonable job fitting 
NaI and CaI, whereas the CO is a little stronger in the
synthetic spectra, suggesting higher temperatures for
11A and 11B. However, we note that at these gravities some over
estimation of the temperatures from the synthetic CO lines were also
seen in the similar case of 2M0141 (Kirkpatrick et al. 2006).

If we let $T_{eff}$ and log(g) be free parameters, the very best fit
to our observed spectra with the synthetic models suggests that
$\Delta T_{eff}\sim200$K between 11A and 11B. Hence, even though there
is some uncertainty in the absolute temperatures derived from these
fits, we can have some confidence in a $\Delta T_{eff}=200$K between
11A and 11B.

\subsubsection{Surface Gravities}

In figure 10 the J-band is compared to synthetic spectra. In the more
 gravity sensitive J-band we find our adopted log(g)=3.95 for Oph 11
 is slightly too low for a good fit to the observed pseudo-continuum. A
 slightly better fit is obtained with log(g)=$4.25\pm0.50$. However,
 the adopted $T_{eff}=2375$K and $T_{eff}=2175$K for 11A and 11B
 combined with log(g)=3.95 produce good agreement on the strength of
 the KI doublets. Hence, it appears that our adopted age of $\sim$5
 Myr (hence log(g)$\sim$3.95) for 11A and 11B is reasonable.  

In summary, the optimal synthetic fits suggest somewhat higher
 gravities (log(g)=$4.25\pm0.50$) but confirm our $\Delta T_{eff}=200$K
 between 11A and 11B. Our synthetic spectra preclude log(g)$<3.75$ and
 so ages $<2$ Myr are very unlikely for Oph 11.

\subsubsection{The SED of 11A and 11B and Their Thermal IR Excess}   

Longward of 3 $\mu$m the SED of 11A and 11B (Table 1; Fig 11) indicates
substantial excess emission compared to our synthetic M9 and M9.5
(log(g)=3.95) spectra for 11A and 11B.  This suggests that both 11A and
11B have circumstellar dust disks. In particular, 11B's excess is very
strong. The discovery by Jayawardhana \& Ivanov (2006b) that both 11A
and 11B have strong (and broad) H$\alpha$ emission suggests active
accretion may be still occurring around both of these objects. While some
$50\pm12$\% of young brown dwarfs have an IR excess suggesting
circumstellar disks, only $\sim16\pm6$\% are estimated to be actively
accreting (Bouy et al. 2006b and references within). {\it Oph \#11 is the first brown dwarf binary where there is strong evidence that both
components are actively accreting}. Similar conclusions were drawn for
the young, single, very low mass, brown dwarf Cha 110913-773444, which also has
an IR-excess (Luhman et al. 2005).
As noted in Section 3.1, extinction toward Oph 11 appears to be small; thus, 
the two dusty disks are not close to edge on.


\subsection{Luminosity and Mass of 11A and 11B}

With an Ophiuchus distance of $125\pm25$ pc (de Geus et al. 1989), an
observed $Ks=13.92\pm0.05$ mag, a K-band bolometric correction of  
$-3.20\pm0.15$ mag 
(Golomoski et al. 2004; appropriate for M9$\pm0.5$), and noting that
for an M9 star $Ks-K\sim 0.03$ mag (Daemgen et al. 2007), we derive a
photospheric luminosity for 11A of $log(L/L_{\sun})=-2.77\pm0.10
$. This luminosity excludes any contribution from the excess
IR emission.  Similarly, we find with $BC_K=-3.10\pm0.15$ mag for an
M$9.5\pm0.5$ that $log(L/L_{\sun}) = -2.96\pm0.10$ for 11B.

Notwithstanding some uncertainty in theoretical evolutionary tracks at
such young ages, and low masses, it is reassuring that the errors
between the DUSTY tracks of Chabrier et al. (2000) and the handful of
dynamical mass calibrated systems are not large when accurate spectra
and K band photometry are known (Luhman \& Potter (2006); Close et al
2007; Stassun et al. 2006). Hence we use our system mass and 
the DUSTY isochrones to estimate masses for 11A and 11B. From the HR
diagram in Figure 12 we see that both 11A and 11B fall close to the 5
Myr isochrone predicted by the DUSTY models. Over the suggested age
range of $3-7$ Myr we find estimated masses from Figure 13 of $13-21
M_{Jup}$ and $10-20 M_{Jup}$ for 11A and 11B, respectively. We are not
the first to caution that these masses, based on theoretical
isochrones at very young ages, may have unknown systematic errors. As
well the range of NIR photometry from Allers et al. (2006),
Jayawardhana \& Ivanov (2006b), 2MASS, and our work suggests 
there may be some variability in the NIR photometry (which would not
be surprising for a system so young).

\subsubsection{Our Masses Compared to those of Jayawardhana \& Ivanov}

As is clear from Table 1, our $\Delta$J, $\Delta$H, and $\Delta$Ks
values are close to that found by Jayawardhana \& Ivanov
(2006b). Hence, we should derive similar masses based on the above
technique which was similar to theirs. However, our absolute
photometric calibration is closer to that of the 2MASS point source
catalog. In particular, our integrated Ks flux is just 6\% brighter
than the 2MASS value, whereas Jayawardhana \& Ivanov (2006b) find
values $\sim$30\% brighter. Their significantly brighter K-band gave
them closer agreement to the more luminous 1 Myr DUSTY isochrone, while we are
closer to the 5 Myr isochrone (see Fig 13).
Whereas, Allers (2005) and
Allers et al. (2006) find Ks=$14.03\pm03$ mag for 11B, some 40\%
brighter than we do in this study. Hence, deriving masses
primarily based on NIR luminosity for \#11 may be problematic for two
reasons: first, there may be an significant underestimation of mass
when one uses J and H fluxes and the dusty models (Close et al. 2005;
2006); secondly, \#11 may be a young variable -- so even the use of
the more reliable K-band luminosity (Close et al. 2005; 2007) may be
misleading. Hence we will also estimate a luminosity (and distance)
{\it independent} mass based on our log(g) and age estimates in the
next section.

\subsubsection{Masses from log(g) and $T_{eff}$}

It is reasonable to assume that the age of 11A and 11B must be $<8$
Myr since there is a strong IR-excess for 11B, strong H$\alpha$
emission (Jayawardhana \& Ivanov (2006b)), and both 11A and 11B are a
reasonable fit to the $\sigma$ Ori 51 brown dwarf whose age $<8$ Myr
(McGovern et al. 2004). However, our fits to the J-band of young, late
M brown dwarf standards suggest that the ages must also be $>2$Myr
(witness the very poor fit of KPNO-Tau4). Moreover, the synthetic
spectral fits find log(g)$>3.75$ also precluding ages $<2$ Myr for
either 11A or 11B. From our age, surface gravity, and temperature
ranges we find masses of $17^{+4}_{-5} M_{Jup}$ and $14^{+3}_{-5}
M_{Jup}$ are appropriate (see Fig. 13). Since these masses avoid some
of the pitfalls of a luminosity based mass estimate we adopt these
mean ($\sim 5$ Myr) values, but we increase the uncertainty ranges to be consistent with our HR
diagram masses of $13-21 M_{Jup}$ and $10-20 M_{Jup}$ for 11A and 11B. Hence our final adopted masses are $17^{+4}_{-5} M_{Jup}$ and $14^{+6}_{-5}
M_{Jup}$. Since we do not include (the
currently unknown) systematic errors of these models, our mass
uncertainties are likely underestimates. These masses are both
slightly above the $\sim$13 $M_{Jup}$ deuterium limit, and above the
14--15 $M_{Jup}$ and 7--8 $M_{Jup}$ masses found by Jayawardhana \&
Ivanov (2006b).

\section {DISCUSSION}

\subsection{Are 11A and 11B Members of the Ophiuchus Cloud Complex?}

Ophiuchus 11B possesses a strong mid-IR (5 \& 8 $\mu m$) excess, and 11A a weaker
one (see Figure 11).  Mid-IR excesses are common only among stars
younger than $\sim$8 Myr (Mamajek et al. 2004; Silverstone et al. 2006;
 Rhee 
et al 2006
and references therein). The strong $H\alpha$ emission measured by
Jayawardhana \& Ivanov (2006b) for 11A and 11B suggests
\#11 is actively accreting and associated with the molecular cloud.
 Indeed, Wilking et al. (2005) find active accreters (CTTSs) ranging
in age from $\sim0.3-10$ Myr in the $r\sim1$pc area around the $\rho$ Oph cloud (including the location of Oph11 \& Oph16).
Our derived $5\pm2$Myr age of \#11 is consistent with Oph 11's
$\sim0.5^{\circ} (\sim1$pc) distance from the $\rho$ Ophiuchus cloud core where
active star formation is taking place. Hence, it is very likely that
11A and 11B are members of the Ophiuchus/Upper Soc sub group of the Sco-Cen OB association. 

Based on the very low extinction ($A_V\sim0$ mag) to Oph\#11 (despite being in the line of sight to the $^{13}CO$ cloud), we suggest that
Oph\#11 is at a distance of $\sim125\pm25$pc. In other words, Oph \#11 is likely slightly
in the front of the $\sim$150 pc ''halo'' of $\sim3$ Myr objects around the
$\rho$ Oph cloud core, which typically have $A_V>1.5$ mag (Wilking et al 2005). To be conservative we adopt distances of 100-150 pc for the system.


\subsection{Do 11A and 11B Form a Common Proper Motion Pair?}

We consider whether 11A might be a foreground dwarf; statistically its
observed proper motion is smaller than the expected proper motion of a
foreground dwarf.
We can use archival POSS II images to determine if the pair's
orientation on the sky is changing (due to 11A being a relatively fast
moving foreground object). The very red optical colors of 11B make it
impossible to detect in the POSS II ``IR'' (RG715) images, only a very
faint object at the current position of 11A is observed, while neither
is seen in the POSS II ``blue'' (GG395 filter).  However, both 11A and
11B appear in the RG610 POSS II ``red'' filter images (Fig, 14) due to
their strong H$\alpha$ emission (Jayawardhana \& Ivanov 2006b).  In
the 1993.33 epoch red POSS II images the binary separation is
$2.05\pm0.20\arcsec$ at $PA=174\pm6^{\circ}$) which may be compared to
our images (Figs. 1 \& 2; separation=$1.943\pm0.022\arcsec$;
PA=$176.2\pm0.5^{\circ}$).

If \#11A were a field M9 dwarf with Ks=13.92 mag, then it should be at
$\sim40\pm5$ pc. Adopting a velocity of $V_{tan}\sim30$ km/s (the
approximate median of late M field dwarfs; Gizis et al. 2000), would
then imply a typical displacement of $\sim2\arcsec$ between 11A and
11B in the epoch 1993.33 DSS images. Since this conflicts (at the
$\sim10\sigma$ level) with what is observed, we conclude that 11A is
not a foreground field dwarf. Hence, 11A and 11B very likely form a
common proper motion pair.

\subsection{Are 11A and 11B Physically Bound?}

Comparison of the epoch 1993.33 and current images shows that the
differential displacement of 11A and 11B over the last 13.22 years is
only $\sim3\pm5$ km/s, consistent with expectations for a physical
system this wide (estimated period $(2\pm1)$x$10^4$ yr; Table 3).
However, since all members of the Ophiuchus association will have
similar proper motions, it is, perhaps, possible that 11A and 11B are
well separated along our line of sight but, by chance, close together
in the plane of the sky.

We can estimate how likely it is to observe two very cool (late
M/early L) objects within $2\arcsec$ of each other in a cloud like Ophiuchus. The density of such objects can be estimated for
the survey of Allers et al. (2006); in 1700 sq. arcmin they found two
$\sim$ M8-M9 objects (\#11 and \#12). Follow-up optical spectra by
Jayawardhana \& Ivanov (2006a) showed that (in the optical) \#12
appears to be a background quasar. We in fact find \#12 to be a
1.3\arcsec ``optical binary'' consisting of a faint z=2 QSO ``12B'' and a slightly
brighter G-giant ``12A''-- from Gemini GNIRS IFU observations). Hence, one would expect in Ophiuchus to find one
late M/early L object in every 1700 sq. arcmin.  Therefore the odds of
a chance alignment of two such objects within $2\arcsec$ is just
$\sim2\times 10^{-6}$. Thus, most probably, 11A and 11B are physically
bound.

Are they still a physically bound pair today?  If they are now
drifting apart at, say, the $\sim$0.5km/s escape velocity of the
system, then the current 237 AU projected separation suggests that
they must have been bound until just $\sim$2x$10^3$ years ago. In
other words, to explain the close proximity of the pair they must have
been bound for $>99$\% of their lives and have only {\it just} become
unbound, a very unlikely scenario. It is, however, possible that the
Oph 11 system will become unbound in the future (see Section 4.6.3.).

\subsection{Our Masses and Spectral Types Compared to those of Other Recent Studies}

Several groups have recently reported spectral types and
masses for Oph 11AB, Allers et al. (2007) find spectral
types of M$7\pm1$ and M$8\pm1$ for 11A and 11B (from the low resolution R$\sim$300 NIR
spectra of Allers 2005). They derived ages of $\sim20$ Myr and masses of 65 
and 35 $M_{jup}$ from dusty models. A more recent work by the same group 
(Luhman et al. 2007) find
types of M7.25 and M8.75 and masses $\sim59 M_{Jup}$
and $\sim21 M_{Jup}$ for 11A and 11B and, like us, they derive an age of
$\sim5$ Myr for the system. 

These two studies derive earlier spectral types than our M$9\pm0.5$ and
M$9.5\pm0.5$ for 11A and 11B. Part of the disagreement is due to a lack of
``gold-standard'' spectral templates at $\sim5$ Myr ages. Hence, both of
these studies use/compose templates of somewhat younger ($\sim1-3$ Myr) ages
(and with different extinction corrections) which can lead to additional
uncertainty distinguishing between age and temperature effects (for example:
cooler older objects have similar alkali line strengths as younger hotter
objects). Moreover, there are systematic errors in the temperature
scale, adding further uncertainty to the final $T_{eff}$.

Very recently Brandeker et
al. (2006b) have also estimated higher masses than Jayawardhana \&
Ivanov (2006b) from new NIR spectra. They confirm our spectral types
and temperatures ($2350\pm150$ K and $2100\pm100$ K for 11A and 11B; 
compared to our 2375 and 2175 K). They derive these temperatures by a pure comparison to the DUSTY models of Baraffe et al. (2002) and the latest brown
dwarf models of Burrows et al. (2006). Over an age range of 1-10 Myr
Brandeker et al. (2007) find masses of $13^{+8}_{-4} M_{Jup}$ and
$10^{+5}_{-4} M_{Jup}$ for 11A and 11B (slightly lower, yet consistent, with our $17^{+4}_{-5} M_{Jup}$ and $14^{+6}_{-5} M_{Jup}$ masses).

In summary, 11A is likely above the deuterium burning limit ($\sim 13 M_{Jup}$)
and the system age is likely closer to 5 Myr than 1 Myr. However, there is
still significant disagreement as to the exact spectral type of 11A with a
range of M7.25-M9 (and dusty model masses of $\sim58-13 M_{Jup}$). In the
case of 11B there is better agreement (M8.75--M9.5; and masses of $21-10
M_{Jup}$). {\it All recent studies agree that Oph 11AB is best described as a
young ($\sim5$ Myr), very wide ($\sim240$ AU), low mass
($M_{tot} \sim23-80 M_{Jup}$) brown dwarf binary}. Moreover, the evidence
presented here indicates that the system is also a bona-fide bound binary 
with
both members likely possessing their own accretion disks.

\subsection{The Oph \#16 System}

We resolved Allers' object \#16 into two components at JHKs (Fig.1 \& Table 
2).  In a similar manner as for Oph \#11 we 
were able to estimate Spitzer [3.6], [4.5], [5.8], and [8.0]
fluxes for both components (Table 2). 

\subsubsection{Extinction, Colors, and Spectral Types for Oph 16A and 16B}

Since Jayawardhana \& Ivanov (2006a) did not obtain optical spectra of
\#16 it is somewhat uncertain what the true spectral types for 16A
and 16B are. We note, however, that 16A and 16B appear significantly
more luminous and bluer (warmer) than 11A or 11B. Moreover, \#16 is
just $0.3^{\circ}$ ($\sim0.6$pc) from the Oph cloud core (D$=125\pm25$ pc),
and well inside the main dark cloud containing the L1688 core. Hence,
Oph 16 would likely have an age of $\sim1$ Myr. If this is true than
we need an extinction of $A_{V}\sim3\pm1$ to have these objects fall
near the 1 Myr isochrone of $\rho$ Oph. This is a reasonable extinction
since Allers et al. (2006) estimated $A_{V}=2$ for this object.

After correcting our NIR photometry for the $A_{V}=3\pm1$ extinction
towards \#16 we find (see Table 2) for 16A J-Ks=$0.99\pm0.31$ which is
consistent with M2-M8 spectral types (Kirkpatrick et al. 1999). 
Hence, we adopt its spectral type as M$5.5\pm3$. 
Future (spatially resolved) spectroscopic observations are
required to obtain better estimates for the spectral types of 16A and
16B \footnote{very recently Allers et al., in prep, observe and confirm these
spectral types}.  However,
we can estimate temperatures for 16A and 16B from our dereddened
NIR colors: For 16A $T_{eff}\sim3000\pm300$K and for 16B 
$T_{eff}\sim2925\pm300$K (Golomoski et al. 2004).  In the same
manner as \#11, we find (with $BC_{K}=-2.95$) $log(L/L_{\sun}) = -1.18\pm0.10$ for 16A and $log(L/L_{\sun}) = -1.47\pm0.10$ for 16B (where
$BC_{K}=-3.00$; Golomoski et al. 2004). The luminosity errors of \#16
take into account our 1 mag of uncertainty in the extinction
correction.

\subsubsection{Are 16A and 16B members of the Cloud Complex?}

From the location of 16A and 16B in the HR diagram of Figure 12 it is
likely that both components are members of the Ophiuchus cloud (consistent with their $\sim$0.6 pc distance from the core). From
the $\sim1$ Myr isochrone (in Figure 12) we estimate masses of
$\sim100 M_{Jup}$ for 16A (likely a VLM star) and $\sim72 M_{Jup}$
for 16B (likely a high mass brown dwarf or VLM star).  Allers et al. (2006)  
estimated a mass for \#16AB of $\sim110 M_{Jup}$.
Spatially resolved spectra in the NIR will be needed to better
constrain the ages of 16A and 16B. However, the strong
mid-IR fluxes of both 16A and 16B in Table 2 suggests that 16A and 16B
are both young VLM members of the Ophiuchus cloud.

\subsubsection{Are 16A and 16B Bound?}

We believe that 16A and 16B form a common proper motion
pair since the orientation of the binary today
(separation=$1.696\pm0.005\arcsec$, PA=$218.29\pm1.00$; Fig. 1) is
consistent with its orientation in 1982.66 (separation=$1.87\pm0.2\arcsec$; PA=$216\pm10^{\circ}$;
 Fig. 14). This is exactly what we would expect for a system with a
very long period of $\sim8$x$10^3$ years (see Table 3).

Is it plausible that 16AB could be a foreground VLM binary? We
estimate (in the same manner as Close et al. 2003) such a hypothetical
foreground system would have a photometric distance of $\sim10$pc to
match 16A's brightness and very red (undereddened) magnitudes
(Ks=10.56 and J-Ks=1.41). Such a VLM binary would have a period of
$\sim170$ years and so in the 23.89 years between Fig. 1 and Fig. 14
one would expect $\sim50^\circ$ of rotation. However, over the last
23.89 years the system has only changed by a $\Delta
PA=1.9\pm10.0^{\circ}$ which shows that Oph 16 is not a
foreground VLM system. Moreover, the IR-excess of 16AB cannot be
explained by a foreground field (old) system. In the same manner as
for Oph \#11 we estimate that the odds of finding two unrelated,
nearly equal magnitude, low-mass, Ophiuchus members within $2\arcsec$
of each other is less than $\sim10^{-5}$. Hence, we conclude that Oph \#16AB is
a newly discovered, wide ($\sim$212 AU), low mass ($M_{total}\sim0.17
M_{\sun}$) VLM binary system in the Ophiuchus cloud complex.

\subsection{Ophiuchus \#11 and \#16 Compared with Other VLM Binaries}

Figures 15 \& 16 show how Oph \#11 and Oph \#16 compare with
other known binary systems. Oph \#11 is the second least
massive binary known (with $M_{tot}\sim0.031 M_{\sun}$; just slightly
more massive than 2M1207) and has the lowest binding energy of any
brown-dwarf binary.  All currently known low binding energy systems
are very young ($<10$ Myr; plotted as open circles in Figures 15 \&
16).  Hence, one might predict that
systems like Oph \#11 will tend to become unbound as
they age.

\subsubsection{How Common Are Very Wide Young VLM Binaries?}

It is interesting to note that very wide, low-mass, binaries may not
be that rare when very young. 
We observed four (not counting \#12) of
the Allers et al. (2006) Oph sample (their \#8, \#10, \#16) and the two M8 objects (GY 264 \&
GY3) of Wilking et al. (2005) with Keck II LGS AO and found \#11 (with NIRI) and \#16 ($\sim33\pm23$\%
of the sample) were binaries, and 100\% of our binaries were wide.

We caution that this is a very small sample, dominated by small number
statistics. Also no correction for the Malmquist bias of the Allers et
al. (2006) sample has been made. Nevertheless, a significant
fraction of the Allers et al. (2006) Ophiuchus sample are young,
wide, VLM binaries. Moreover, Bouy et al. (2006) find in their NGS AO
survey of slightly more massive binaries in the nearby Upper Sco
region that 3/9 of their binaries were wide VLM systems.  One of
these binaries, Denis PJ161833.2-251750.4AB, is spectroscopically
confirmed (Luhman 2005) and another, USCO 1600AB, is likely also bound (Bouy
et al. 2006). In contrast HST/ACS surveys of Upper Sco (5 Myr; Kraus
et al. 2005) found no young wide VLM binaries. Bouy et al. estimate
(including the non-detection of Kraus et al. 2005) that (3/12)
$\sim25\%$ of VLM binaries in Upper Sco are wide. In summary, Bouy et al. find two
solid wide ($>100$ AU) VLM binaries out of 40 VLM
objects; hence $f_{VLM_{wide\&young}}\sim 5^{+6}_{-2}\%$ in the Upper
Sco OB association.

The ($\ga4$ AU) spatial resolution survey of the lower density
Taurus association of Kraus et al. (2006) with HST/ACS found 2/22 VLM
objects were binary, but none were wider than $\sim6$ AU. Combining
our (2/6) result and the (0/22) result by Kraus et al. (2006) suggests
that 2/28 VLM objects are wide binaries. Hence,
$f_{VLM_{wide\&young}}\sim 7\pm5\%$ in Taurus and Ophiuchus. This is
very consistent with the $5^{+6}_{-2}\%$ found in the Upper Sco
cluster. Merging all the surveys indicates 4/68 young ($<10$ Myr) VLM
objects that were observed at high resolution ($sep\ga 5$ AU) were
found to be wide ($>100$ AU) binaries. 
Therefore, we adopt
$f_{VLM_{wide\&young}}\sim 6\pm3\%$. While even larger surveys with LGS AO
and HST will be required to confirm these frequencies, {\it it is
clear that a significant ($\sim6\%$) fraction of young VLM objects are
formed in wide ($>100$ AU) binaries.}

\subsubsection{Is There a Dearth of Wide VLM Binaries in the Field?}


In this section we try to estimate the frequency of wide VLM binaries
in the field ($f_{VLM_{wide\&old}}$) based on published surveys. After
correcting for the selection effect that field objects are fainter than
young objects of a given mass, we can derive the depletion of old
field wide systems to that of young systems. We define $X_{evap}\tbond
f_{VLM_{wide\&young}}/ f_{VLM_{wide\&old}}$, as the fraction of wide
VLM binaries that have ``evaporated'' as these binaries age.

Of the 69 field (old; 0.5--10 Gyr) VLM ($M_{tot}<0.2 M_{\sun} $)
binaries currently known 
(VLM Binary Archive 
\footnote{http://paperclip.as.arizona.edu/ $\sim $ nsiegler/VLM\_binaries \\ Maintained by Nick Siegler}
) only Denis 0551 is wider than 100
AU. Hence only $\sim 1.4\%$ of known field VLM binaries are wide. All
searches to find wide VLM binaries (Close et al. 2003; Gizis et
al. 2003; Bouy et al. 2003; Burgasser et al. 2003; Siegler et
al. 2005; Allen et al. 2007; Reid et al. 2006; Burgasser et al. 2006
and references within) have all failed to find a single wide ($>50$
AU) VLM field binary despite being sensitive to $sep\ga2$ AU and
masses$\ga 13 M_{jup}$. The 132 VLM (M7-L8) study of Allen et
al. (2007) alone suggests the frequency of wide VLM systems must be
$f_{VLM_{wide\&old}}<2.3\%$ in the field at the 95\% confidence level.

Only the seeing limited survey of Billeres et al. (2005) found one
wide system: Denis 0551. While it is difficult to estimate 
the total
number of unique VLM objects searched we note that Close et al. 2003
\& Siegler
et al. imaged 80 M6-L0 systems with AO; Reid et al. 2006 looked at
52 L dwarfs and Burgasser et al. 2003 looked at 10 T dwarfs with HST; Allen
et al (2007) has looked at 123 M7-L8 systems. Hence out of $\sim265$
unique VLM objects (this is a lower limit since not all studies publish their null results), none were in wide binaries. Hence we can estimate an upper limit to
$f_{VLM_{wide\&old}}< 1/265=0.37\%$. The lower limit to frequency can
be bounded by the fact that one wide system exists, hence of the 580 L
\& T field dwarfs known (Dwarf Archives
\footnote{http://spider.ipac.caltech.edu/staff/davy/ARCHIVE/ \\
Maintained by Davy Kirkpatrick, Chris Gelino, \& Adam Burgasser}), only
one is a wide binary, so $f_{VLM_{wide\&old}}> 1/580=0.17\%$. Hence we
adopt $f_{VLM_{wide\&old}}\sim0.3\pm0.1\%$. This is much smaller than
our young wide VLM frequency of $6\pm3\%$. While a more careful
analysis of the problem is warranted, {\it it is still clear that the
frequency of wide ($>100$ AU) VLM binaries ($M_{tot}<0.2 M_{\sun}$) is
$\sim20^{+25}_{-13}$ times higher for young ($<10$ Myr) VLM objects than
for old field VLM objects.}

A small observational selection effect is that a much older ($\sim$5
Gyr) analog of Oph\#11B maybe too cool (very late T or ``Y'' spectral type)
and too faint to be easily detected in the field today. However, older
Oph\#16AB, Denis 1618AB, and USCO 1600AB systems
 in the
field would be detected by the 2MASS survey as L dwarf primaries with
L or early T dwarf secondaries (detectable by the aforementioned VLM
binary surveys). Correcting for this selection effect (throwing out Oph\#11) yields a slightly smaller ``field
detectable'' wide VLM binarity of $3/68=4.4\pm2.5\%$ and hence a corrected
$X_{evap}=15^{+20}_{-10}$. The lack of old wide systems is real, it is not a selection effect of our wide young binary population being too low in mass to be detected in the older field. 

\subsubsection{Do Wide VLM Binaries Evaporate as they Age?}

One amelioration of the low ($\sim0.3\pm0.1\%$) wide binary frequency in the
field and the higher frequency of young ($<10$ Myr)
wide VLM binaries is that these wide VLM systems are dynamically dissipating, or evaporating,
over time. After all, a differential ``kick'' of order
$V_{esc}\sim1$ km/sec would be more than adequate to dissolve all the
wide VLM binaries in Figure 15.

Why would such a tidal kick occur? Over the lifetime of a wide binary there
will be many smaller stochastic encounters (with other stars or molecular
clouds; Weinberg et al. 1987), which can increase the separation of a wide binary slowly over
time. Eventually, these encounters may also disrupt the binary.

However, it is not certain by which mechanism VLM binaries will become
unbound.  Separations of $\sim200$ AU are still very small
compared to the average $10^{5-6}$ AU separations between stars in the
galaxy (whereas wide {\it stellar mass} binaries have separations of
$\sim10^4$ AU, and so are more easily disrupted). Hence, the odds of
any one encounter being close enough ($\la 240$ AU) to tidally dissolve
a wide (200 AU) VLM binary are very low (Close et al. 2003; Burgasser
et al. 2003) unless the stellar density is high.

To investigate the stability of wide binaries we note that Weinberg et al.(1987)'s analytic
solution of Fokker-Planck (FP) coefficients describing advective diffusion
of a binary due to stellar encounters is $t_{*}(a_{o})\sim
3.6\times10^5(n_{*}/0.05
pc^{-3})^{-1}*(M_{tot}/M_{\sun})*(M_{*}/M_{\sun})^{-2}*(V_{rel}/20kms^{-1})(a_{o}/AU)^{-1}
$Gyr where $t_{*}(a_{o})$ is the time required to evaporate a binary
of an initial semi-major axis of $a_o$, the number
density of stellar perturbers is $n_{*}$ of mass $M_{*}$ and relative velocity
$V_{rel}$ (adopted from Weinberg (1987); assuming, as they do, that
their $ln\Lambda \sim 1$). Hence, the {\it maximum} projected
separation of a bound binary (assuming semi-major axis $a= 1.26 \times sep$;
Fischer \& Marcy 1992) after 10 Gyr in the field is
given by:

\begin{equation}
{sep^{diffusive*}_{field} \la 28\times10^3\left(\frac{0.16}{0.05 pc^{-3}}\right)^{-1}\left(\frac{M_{tot}}{M_{\sun}}\right)\left(\frac{0.7}{M_{\sun}}\right)^{-2} \sim 1800\left(\frac{M_{tot}}{0.1 M_{\sun}}\right) AU}
\end{equation}

where we have used the measured Galactic disk mass density of $0.11 M_{\sun}/pc^{-3}$
and an average perturber mass of $0.7 M_{\sun}$, and $V_{rel}\sim20$
km/s (Pham et al. 1997; Holmberg \& Flynn 2000).

In addition to the evaporation of binaries due to diffusion there is
also the chance of a catastrophic encounter evaporating the
binary. While such encounters are less important than diffusion, they cannot be completely ignored. From the work of
Weinberg et al. we find the {\it maximum} projected separation ($sep$) of a
binary to stay bound w.r.t. close encounters over 10 Gyr in the field is:

\begin{equation}
{sep^{catastrophic*}_{field} \la 52\times10^3\left(\frac{0.16}{0.05 pc^{-3}}\right)^{-1}\left(\frac{M_{tot}}{M_{\sun}}\right)\left(\frac{0.7}{M_{\sun}}\right)^{-2} \sim 3300\left(\frac{M_{tot}}{0.1 M_{\sun}}\right) AU}
\end{equation}

It is clear that all field VLM binaries with $M_{tot}\ga 0.1
M_{\sun}$ will be stable against any type of stellar encounter
as long as $sep\la 1800$ AU. To better illustrate these regions of
stability we plot the above two relations on the right-hand side of Figure 17. Note how in Fig. 17 all
the VLM binaries are not in the ``Field Unstable'' region
(since $sep<<sep^{diffusive*}_{field}<sep^{catastrophic*}_{field}$). Therefore, we can conclude that no known VLM binary will be
evaporated due to random stellar encounters in the field. However,
these limits do help us understand the distribution of wide stellar
mass binaries like those of Close et al. 1990 (solid triangles in
Figure 17). Indeed no stellar binaries are observed in the ``Field
Unstable'' region as one would expect.

This still leaves us with the problem of why $X_{evap}$ is $\sim
15^{+20}_{-10}$. Besides being young, all the known wide VLM systems
are in the Chamaeleon I (2M1101), or Ophiuchus/Upper Sco/Sco-Cen OB
(Oph 11, Oph 16, Denis 1618) clusters. Since 2M1101 is near the lower
core of Chamaeleon I, with $\sim100$ members with a core radius of
$\sim0.25$pc (Luhman 2004) we estimate that near 2M1101 $n_{*}\sim
1500/pc^{3}$. For the Ophiuchus core there are $>200$ members within a
radius of $\sim0.3$ pc (Wilking et al. 2005) and so $n_{*}\sim
1800/pc^{3}$. At the distance of Oph\#11 and Oph\#16 there are
$\sim100$ more members but the density drops to $n_{*}\sim
50-220/pc^{3}$ (Wilking et al. 2005). In summary, we adopt a mean
$n_{*}\sim 1000/pc^{3}$ for these ``clusters'' (in general agreement
with other similar sized clusters; Gutermuth et al. 2005). These ``clusters'' (or associations) are
much higher densities than the field. It is also true that the
encounters have much longer interactions timescales (since $V_{rel}\la
3$ km/s), hence the clusters where these stars formed will play a role
in evaporating them before they can join the field. If we assume that
these objects are subjected to an additional $\sim10$ Myr of the mean
cluster density then from the above equations we have:

\begin{equation}
{sep^{diffusive*}_{cluster} \la 28\times10^{3}\left(\frac{1000}{0.05 pc^{-3}}\right)^{-1}\left(\frac{M_{tot}}{M_{\sun}}\right)\left(\frac{0.7}{M_{\sun}}\right)^{-2}\left(\frac{3}{20 km/s}\right) \sim 44\left(\frac{M_{tot}}{0.1 M_{\sun}}\right) AU}
\end{equation}

In the case of a catastrophic encounter we have:

\begin{equation}
{sep^{catastrophic*}_{cluster} \la 52\times10^3\left(\frac{1000}{0.05 pc^{-3}}\right)^{-1}\left(\frac{M_{tot}}{M_{\sun}}\right)\left(\frac{0.7}{M_{\sun}}\right)^{-2}\left(\frac{3}{20 km/s}\right) \sim 80\left(\frac{M_{tot}}{0.1 M_{\sun}}\right) AU}
\end{equation}

Hence, after just 10 Myr the cluster has a much more disruptive effect
on the wide VLM binaries than do encounters in the field over 10
Gyr. From Figure 17 we see that all known wide ($sep>100$ AU) young
($<10$ Myr) VLM binaries are found in the ``Cluster Unstable'' region
where $sep^{diffusive*}_{cluster}\la sep \la
sep^{diffusive*}_{field}$. Therefore, there is a good chance that
all the known wide, young, VLM binaries are in the process of
evaporating in their clusters. Indeed past cluster member
(non-catastrophic) encounters may have already played a role in
creating the current (very wide) separations observed for this handful of objects. 

Our simple analysis above is useful but approximate. However, an
independent analytic solution of this problem by Ivanova et al. (2005)
and Brandeker et al. (2006) implies that a $M_{tot}<0.1M_{\sun}$
binary is ``soft'' if $a>64$ AU. Hence, all our wide, young, VLM systems can be
evaporated by a strong encounter. The timescale for such an encounter
by their equations is larger than found above (our equation 4), yet
their solutions still predict that Oph 11 would become unbound in $\sim11$
Myr similar to our expectations above. 

Of course none the analysis above implies that all these binaries must
evaporate, only that it is possible. A more detailed numerical analysis
of each system, in each cluster, would have to be carried out to
ascertain individual outcomes. Indeed some of these may escape their
cluster before evaporation (as may have been the case for Denis
0551). 

If we assume that $\sim10\%$ of
VLM objects form in very low mass groups ($n_{*}\la10/pc^{3}$), then their wide systems will likely survive (whereas none survive
for the 90\% formed in clusters). Then roughly $\sim6\pm3\%$ of these isolated VLM objects should be in
wide systems. Hence, one expects $\sim0.6\pm0.3\%$ of the field to be
composed of wide VLM systems. Since this is consistent with our
observed value of $f_{VLM_{wide\&old}}\sim0.3\pm0.1\%$ we can take some comfort that the above argument might naturally explain why $X_{evap}$ is observed to be $15^{+20}_{-10}$.

Our scenario does predict that once the system's birth clusters
themselves dissolve that $F_{VLM_{wide}} \sim F_{VLM_{wide\&old}}$. In
other words, once the VLM binaries leave their clusters, the fraction
of wide VLM binaries is ``frozen'' at the $F_{VLM_{wide\&old}}\sim
0.3\pm0.1\%$ level. We can test this prediction by looking in low
density open clusters of intermediate ages to test if $F_{VLM_{wide}}
\sim 0.3\pm0.1\%$.  The ideal cluster for such a search would be the
Pleiades (age $\sim$120 Myr; d$\sim$120 pc). Several large HST \& AO surveys have
failed to detect a single wide VLM binary in this cluster (Martin et
al. 2003; Bouy et al. 2006c). A similar lack of wide system have been found in the older Hyades
(Siegler et al. 2003). So by $\sim100$ Myr there is some evidence that
the wide binary population has mainly been evaporated, as Figure 17
and equations 3 \& 4 would predict.

For completeness we note that disruption by molecular clouds does not
likely play a significant role, since the critical FP impact parameter
$b^{GMC}_{FP}\sim 1\times10^5$AU which is too wide to effect $\sim200$ AU VLM
binaries. However, the much wider ($\sim 2\times10^4$ AU) stellar mass
binaries are certainly affected by molecular clouds and their $R\sim
2-3$ pc sub-clumps (Weinberg 1987), but they are not relevant in the
VLM regime.

\section{Conclusions}

We have obtained the first
spatially resolved $R\sim2000$ near-infrared (J \& K) spectra, mid-IR
photometry, and orbital motion estimates for the wide brown dwarf binary Oph \#11 (Oph 1622-2405). We estimate for 11A and 11B
gravities (log(g)$>3.75$), ages ($5\pm2$ Myr), luminosities
(log(L/$L_{\sun}$)=$-2.77\pm0.10$ and $-2.96\pm0.10$), and
temperatures ($T_{eff}=2375\pm175$ and $2175\pm175$ K). We find
self-consistent DUSTY evolutionary model (Chabrier et al. 2000) masses
of $17^{+4}_{-5} M_{Jup}$ and $14^{+6}_{-5} M_{Jup}$, for 11A and 11B
respectively. Our masses are higher than the previously reported
13--15 $M_{Jup}$ and 7--8 $M_{Jup}$ masses of Jayawardhana \& Ivanov
(2006b). {\it Hence, we find the system is unlikely a ``planetary mass
binary'', but it has the second lowest mass and lowest binding energy
of any known binary}.

 The \#16 binary (Oph 1623-2402) is also an unusually wide (projected
separation of 212 AU), very low-mass (VLM), binary composed of a
$\sim100 M_{Jup}$ primary (16A) and a $\sim73 M_{Jup}$ (16B)
secondary.

While young VLM datasets are dominated by small number
statistics, they are suggestive of a moderately common ($6\pm3\%$)
young ($<10$ Myr) population of wide ($>100$ AU), very low-mass
($M_{tot}<0.2M_{\sun}$), binary systems like Denis PJ161833.2
-251750.4, 2MASS J11011926-7732383, Oph\#11 (Oph 1622-2405), and
Oph\#16 (Oph 1623-2402). Our estimate of the wide young
binary fraction agrees well with the $5^{+6}_{-2}\%$ found in the
Upper Sco by Bouy et al. (2006). We argue that, over
time, the majority of these wide VLM systems become unbound leaving
just the surviving (most tightly bound) fraction observable in the
field today.

We suspect these wide systems likely become unbound since only $\sim
0.3\pm0.1\%$ of field VLM objects are wider than 100 AU today. Hence
it appears that the field population is depleted in wide VLM binaries
by a (selection effect corrected) factor of $X_{evap} =
15^{+20}_{-10}$. The exact mechanism for this evaporation of wide VLM
systems is unlikely to be interactions with stars in the field or with
molecular clouds and their sub-clumps. However, from the Fokker-Planck
solutions of Weinberg et al. (1987), we find stellar encounters with
``birth cluster stars'' may be efficient in dissolving these wide binaries.
Indeed such encounters may help boost the observed $sep$ past the
$sep^{catastrophic*}_{cluster} \sim 80\left(\frac{M_{tot}}{0.1
    M_{\sun}}\right)$ AU limit where catastrophic cluster encounters
  can occur. More detailed numerical studies will be required to
  better constrain the evolution of this evaporating population.

\acknowledgements

We thank NASA for providing observing time on Keck for this
program. We thank both the directors of Gemini and Keck for allocating
DDT time for follow-up observations for this project. We thank Mark
McGovern, Davey Kirkpatrick, and Gael Chauvin for sharing electronic versions of their
spectra with us and Ian McLean for assistance with obtaining the NIRSPEC 
data. We thank Adam Burgasser for helpful discussions. LMC is supported by an NSF CAREER award and the NASA
Origins of Solar Systems program.

CM and BM note that their research was performed under the auspices of
the US Department of Energy by the University of California, Lawrence
Livermore National Laboratory under contract W-7405-ENG-48, and also
supported in part by the National Science Foundation Science and
Technology Center for Adaptive Optics, managed by the University of
California at Santa Cruz under cooperative agreement AST 98-76783.

 These results were based on observations obtained at the Gemini
Observatory, which is operated by the Association of Universities for
Research in Astronomy, Inc., under a cooperative agreement with the
NSF on behalf of the Gemini partnership: the National Science
Foundation (United States), the Particle Physics and Astronomy
Research Council (United Kingdom), the National Research Council
(Canada), CONICYT (Chile), the Australian Research Council
(Australia), CNPq (Brazil) and CONICET (Argentina).

This publication makes use of data products from the Two Micron All
Sky Survey, which is a joint project of the University of
Massachusetts and the Infrared Processing and Analysis
Center/California Institute of Technology, funded by the National
Aeronautics and Space Administration and the National Science
Foundation.





\clearpage
\begin{deluxetable}{lllllllllll}
\tabletypesize{\scriptsize}
\tablecaption{The Oph \#11 Brown Dwarf Binary (Separation=$1.943\pm0.022\arcsec$; PA=$176.2\pm0.5^{\circ}$)\label{tbl-1}}
\tablewidth{0pt}
\tablehead{
\colhead{Parameter} &
\colhead{J} &
\colhead{H} &
\colhead{Ks} &
\colhead{Ls} &
\colhead{[3.6]}&
\colhead{[4.5]}&
\colhead{[5.8]}&
\colhead{[8.0]}&
}
\startdata
$\Delta$ mag & $0.82\pm0.03$ & $0.69\pm0.03$ & $0.52\pm0.03$ & $0.36\pm0.04$ & $0.24\pm0.04$ & $0.11\pm0.04$ & $-0.20\pm0.05$ & $-0.51\pm0.10$\\ 
11A (mag)\tablenotemark{a} & $15.04\pm0.05$ & $14.19\pm0.07$ & $13.92\pm0.07$ & -- & $13.24\pm0.04$ & $13.08\pm0.03$ & $12.96\pm0.05$ & $12.84\pm0.11$ \\
11B (mag)\tablenotemark{a} & $15.86\pm0.05$ & $14.88\pm0.05$ & $14.44\pm0.44$ & -- & $13.48\pm0.04$ & $13.19\pm0.04$ & $12.76\pm0.05$ & $12.34\pm0.08$ \\
\enddata
\tablenotetext{a}{the J, H, and Ks fluxes are standard MKO magnitudes determined at the Gemini telescope. The Spitzer magnitudes at [3.6], [4.5], [5.8], and [8.0] $\mu$m are based on the standard Vega IRAC zeropoints of 276.79, 179.5, 116.69, and 63.122 Jy respectively. We determined new IRAC photometry with a proper binary PSF fit to the IRAC images. There may be an additional $\sim$15\% absolute calibration uncertainty in the IRAC photometry (Evans et al. 2003).}
\end{deluxetable}

\begin{deluxetable}{llllllllll}
\tabletypesize{\scriptsize}
\tablecaption{The Oph \#16 VLM Binary (Separation=$1.696\pm0.005\arcsec$; PA=$218.29\pm1.00^{\circ}$)\label{tbl-2}}
\tablewidth{0pt}
\tablehead{
\colhead{Parameter} &
\colhead{J} &
\colhead{H} &
\colhead{Ks} &
\colhead{[3.6]}&
\colhead{[4.5]}&
\colhead{[5.8]}&
\colhead{[8.0]}&
\colhead{[24] (mJy)}&
}
\startdata
$\Delta$ mag & $0.691\pm0.030$ & $0.691\pm0.022$ & $0.757\pm0.029$ 
& $0.65\pm0.03$ & $0.67\pm0.03$ & $0.75\pm0.04$ & $0.75\pm0.04$ & --\tablenotemark{a}\\
16A (mag)\tablenotemark{b} & $11.19\pm0.26$ & $10.54\pm0.17$ & $10.20\pm0.12$
& $9.88\pm0.03$ & $9.60\pm0.03$ & $9.10\pm0.03$ &$8.12\pm0.16 $  &  $32\pm5$\tablenotemark{a}\\
16B (mag)\tablenotemark{b} & $11.94\pm0.26$ & $11.23\pm0.17$ & $10.88\pm0.12$ 
& $10.53\pm0.06$ & $10.27\pm0.05$ & $9.85\pm0.04$ & $9.31\pm0.10$ & $16\pm3$\tablenotemark{a} \\
\enddata
\tablenotetext{a}{$\Delta$[24] was not determined, however, since 16B is consistently $\sim$0.75 mag fainter we have estimated that this ratio holds for the[24] $\mu$m mJy flux as well.}
\tablenotetext{b}{Above magnitudes were dereddened by $A_{V}=3\pm1$ mag (similar to the $A_{V}=2$ of Allers et al. 2006). The error in the extinction dominates the J, H, and Ks photometric errors. With the $R_{V}=3.1$ extinction law of Fitzpatrick (1999) we have dereddened our observed Oph 16A \& B magnitudes by 0.78 mag at J, 0.51 at H and 0.36 at Ks for $A_{V}=3$. Our integrated Oph 16 flux calibration is based on integrated flux values from the 2MASS point source catalog and the integrated IRAC values are from Allers et al. (2006).}
\end{deluxetable}

\begin{deluxetable}{llllllllll}
\tabletypesize{\scriptsize}	
\tablecaption{The Young, Wide, Low-Mass Ophiuchus Binaries (D=$125\pm25$ pc)\label{tbl-3}}
\tablewidth{0pt}
\tablehead{
\colhead{Allers}&
\colhead{2MASS}&
\colhead{SpT}&
\colhead{$T_{eff}$}&
\colhead{Lum.}&
\colhead{Gravity\tablenotemark{a}}&
\colhead{Age}&
\colhead{Mass} &
\colhead{Pro. Sep.}&
\colhead{Period\tablenotemark{b}} \\
\colhead{\#}&
\colhead{System Name}&
\colhead{}&
\colhead{(K)}&
\colhead{log($L_{\sun}$)}&
\colhead{log(g)}&
\colhead{(Myr)}&
\colhead{(Jup)} &
\colhead{(AU)}&
\colhead{(x$10^{3}$ yr)}
}
\startdata
11A & 16222521-2405139 & M$9\pm0.5$ & $2375\pm175$ & 
 $-2.77\pm0.10$ & $4.25\pm0.50$ & $5\pm2$ & $17.5\pm2.5$\tablenotemark{c} & $243\pm55$ & $20\pm10$\\
11B & -- & M$9.5\pm0.5$ & $2175\pm175$ & 
 $-2.96\pm0.10$ & $4.25\pm0.50$ & -- & $15.5\pm2.5$\tablenotemark{c} & -- & -- \\
16A & 16233609-2402209 & M$5\pm3$\tablenotemark{d} & $3000\pm300$ & 
 $-1.18\pm0.10$ & -- & $\sim1$ & $\sim100$\tablenotemark{e} & $212\pm43$ & $\sim8$\\
16B & -- & M$5.5\pm3$\tablenotemark{d} & $2925\pm300$ & 
 $-1.47\pm0.10$ & -- & -- & $\sim73$\tablenotemark{e} & -- & --\\
\enddata
\tablenotetext{a}{Surface gravities estimated based on the best synthestic spectra fits to our J and K spectra. However, a gravity of log(g)$\sim3.95$ corresponds to the best fit of Oph\#11A and Oph\#11B to the spectral standards.}
\tablenotetext{b}{Periods estimated based on face-on circular orbits.}
\tablenotetext{c}{Masses of 11A and 11B determined from the full range of masses consistent with all of our determined range of gravities, ages, luminosities, and temperatures from the DUSTY tracks (Chabrier et al. 2000). No additional error due to unknown systematics in the models themselves have been added, hence these mass errors are likely underestimated.}
\tablenotetext{d}{Since no spectra have been obtained for 16A and 16B, these spectral types are simply estimated from our dereddened NIR colors from Table 2.}
\tablenotetext{e}{Masses of 16A and 16B estimated from the 1 Myr isochrone. Without a detailed spectroscopic observations it is impossible to be more accurate about the ages or masses of Oph 16 at this time.}
\end{deluxetable}

\clearpage

%

\begin{figure}
 \includegraphics[angle=0,width=\columnwidth]{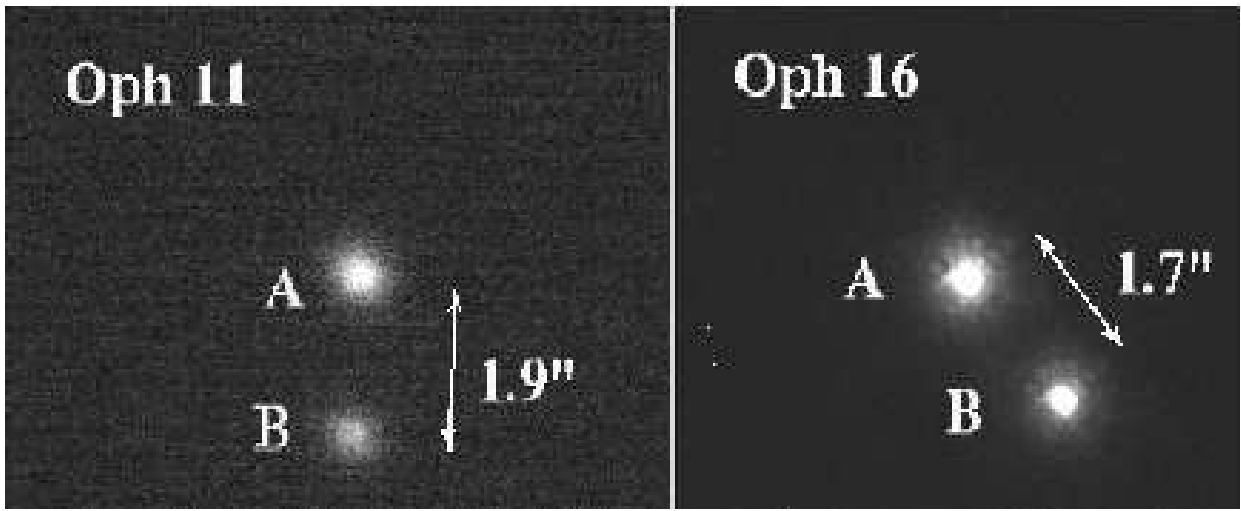}
\caption{
Images of both binary systems found from our survey of Ophiuchus VLM objects. All images are in the Ks
band. The Oph \#11 image is seeing limited (FWHM=$0.3\arcsec$) with
the NIRI IR camera ($0.02177\arcsec pix^{-1}$) at the Gemini North
Telescope. Whereas images of Oph \#16 were obtained at the Keck II
telescope with NIRC2 camera ($0.010\arcsec pix^{-1}$) and the LGS AO
system. Locking on Oph \#16 as its own tip-tilt star gave excellent
$FWHM=0.05\arcsec$ LGS AO (Wizinowich et al 2006) corrected images. North is 
up and East left
in all images.  }
\label{fig1}
\end{figure}

\begin{figure}
 \includegraphics[angle=0,width=\columnwidth]{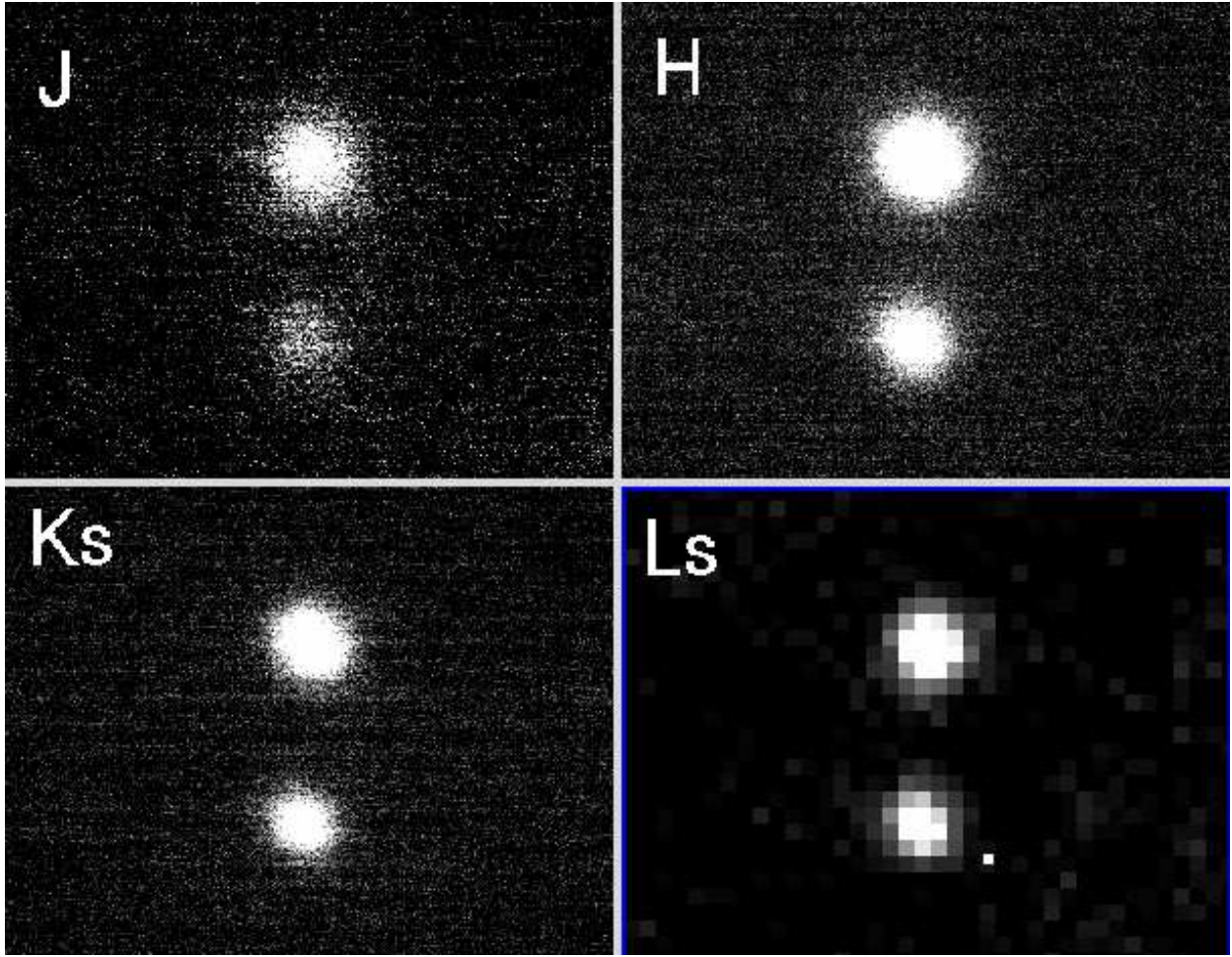}
\caption{
Images of the Oph \#11 system at J, H, \& Ks (obtained in excellent
$FWHM\sim0.3\arcsec$ seeing at Ks) with the NIRI IR camera at the Gemini
North Telescope. The Ls image is from the NIRC IR camera ($0.15\arcsec pix^{-1}$) at the
Keck I telescope. North is up, east to the left.  }
\label{fig2}
\end{figure}

\begin{figure}
 \includegraphics[angle=0,width=\columnwidth]{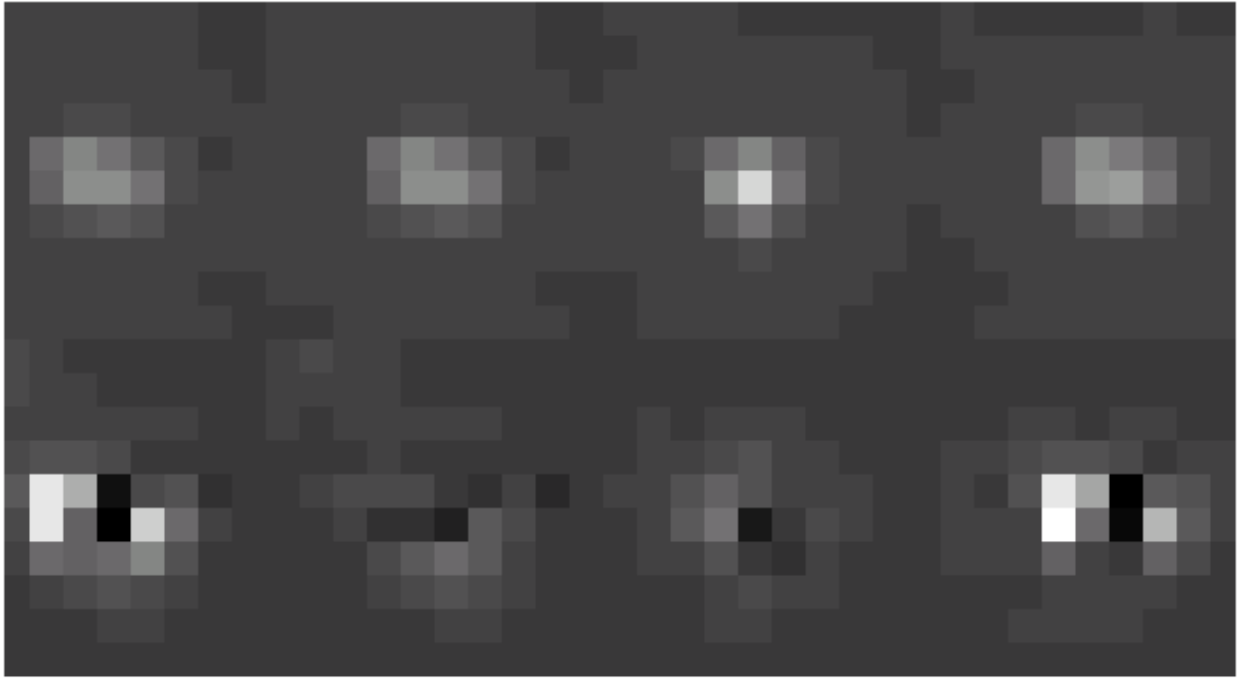}
\caption{
IRAC [3.6] micron ($1.2\arcsec pix^{-1}$) image of \#11 (upper left; note
north is $80^{\circ}$ counterclockwise from up). In the lower left we
show the residual (stretch intensified by 5 times for clarity) of the
system fit to 1 PSF (clearly Oph \#11 is not well fit by a single
PSF). The second column from the left shows our \#11 binary model
($Sep=1.94\arcsec$, $PA=176^{\circ}$) with the best fit fluxes (in this
case with a $\Delta$[3.6] of 0.24 mag). The next ($3^{rd}$) column
from the left shows a single star and its residual when fit with our
IRAC PSF (giving a similarly good residual as our binary model). The upper 
image in the
right-hand column shows a simulated system similar to the \#11AB binary  
and then
below it the residuals left after fitting a single PSF fit (note the 
similarity of this simulation to the comparable fit to the actual data 
shown in the left-hand column). The robustness of the binary fit gives
confidence in our IRAC photometry to the $\sim$0.03-0.05 mag
level. Similar fits were also obtained at [4.5], [5.8], and [8.0] $\mu$m
for \#11 and \#16; see Tables 1 \& 2 for the resulting mid-IR
photometry.  }
\label{fig3}
\end{figure}

\begin{figure}
 \includegraphics[angle=0,width=\columnwidth]{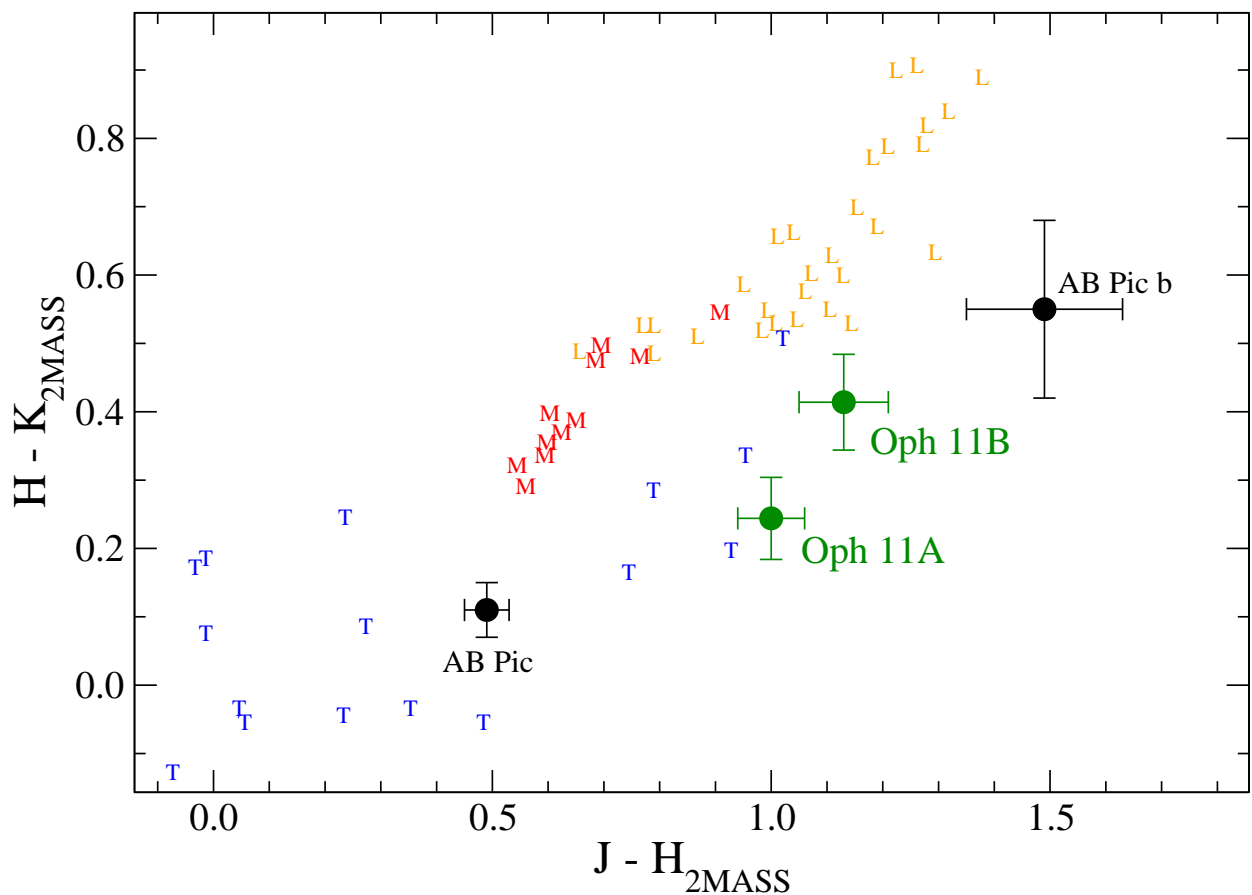}
\caption{
The J-H, H-Ks color-color diagram for field (old) low-mass M, L, \& T
dwarfs. Note how the positions of 11A and 11B are similar to that of
AB Pic b another very low-mass, young, object (Chauvin et al. 2005b). The 
offset in the
positions of 11A and 11B from the dwarf locus is likely due to the low
surface gravity of these objects compared to field dwarfs (Song et
al. 2006). Figure adopted from Song et al. (2006).  }
\label{fig4}
\end{figure}

\begin{figure}
 \includegraphics[angle=0,width=\columnwidth]{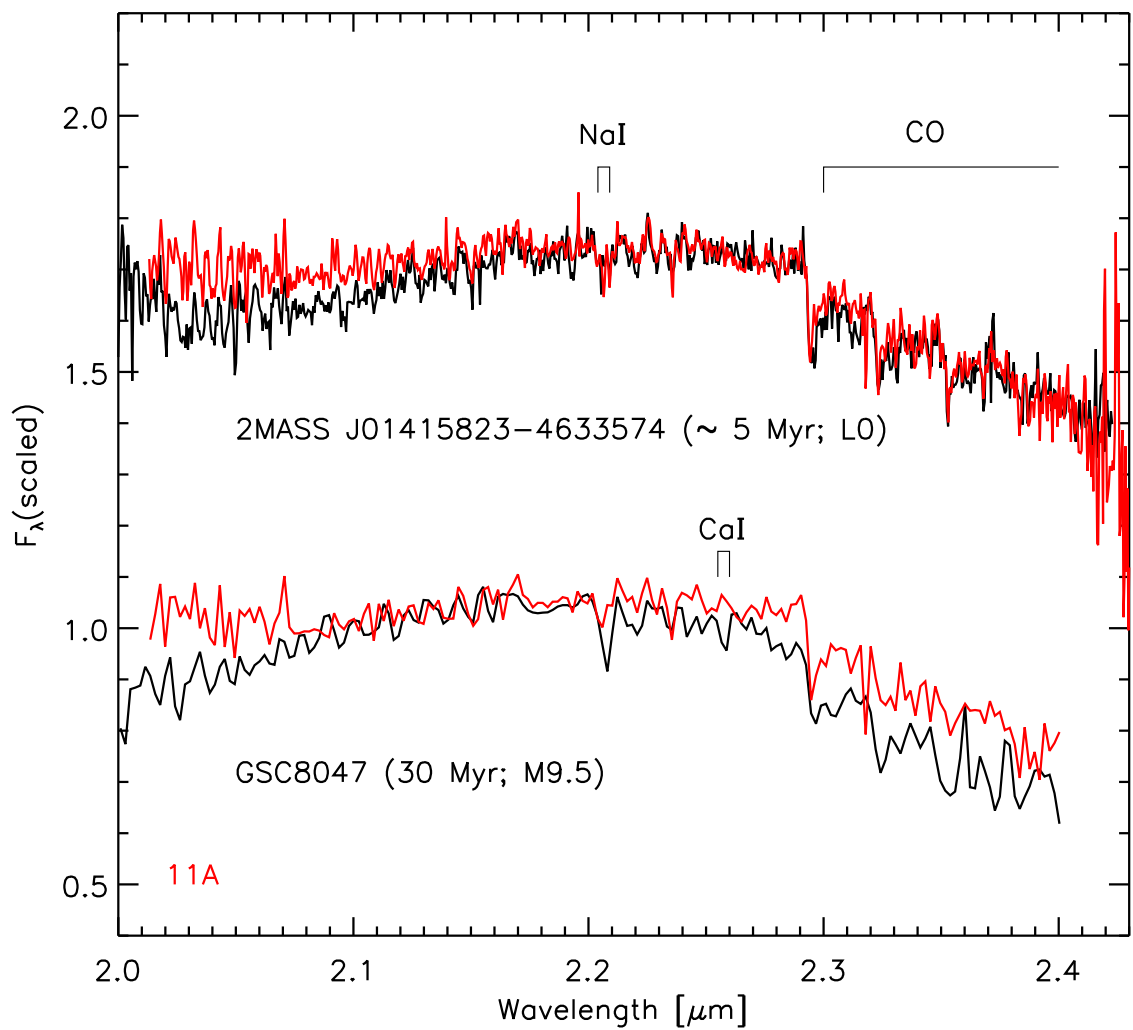}
\caption{
The K-band NIRSPEC spectrum of Oph 11A (red line). Note the good
 agreement between the young L0 dwarf 2M0141 (from Kirkpatrick et
 al. 2006) and 11A. In particular, the good fit to the temperature
 sensitive CO lines suggests 11A is just slightly warmer than a young
 L0. The poorer fit of the gravity sensitive NaI and CaI to the older
 (30 Myr) star GSC8047 (from Chauvin et al. 2005a) suggests that 11A
 must be younger than 30 Myr (we have lowered the resolution of the Oph 11A spectra to match that of GSC8047).}
\label{fig5}
\end{figure}

\begin{figure}
 \includegraphics[angle=0,width=\columnwidth]{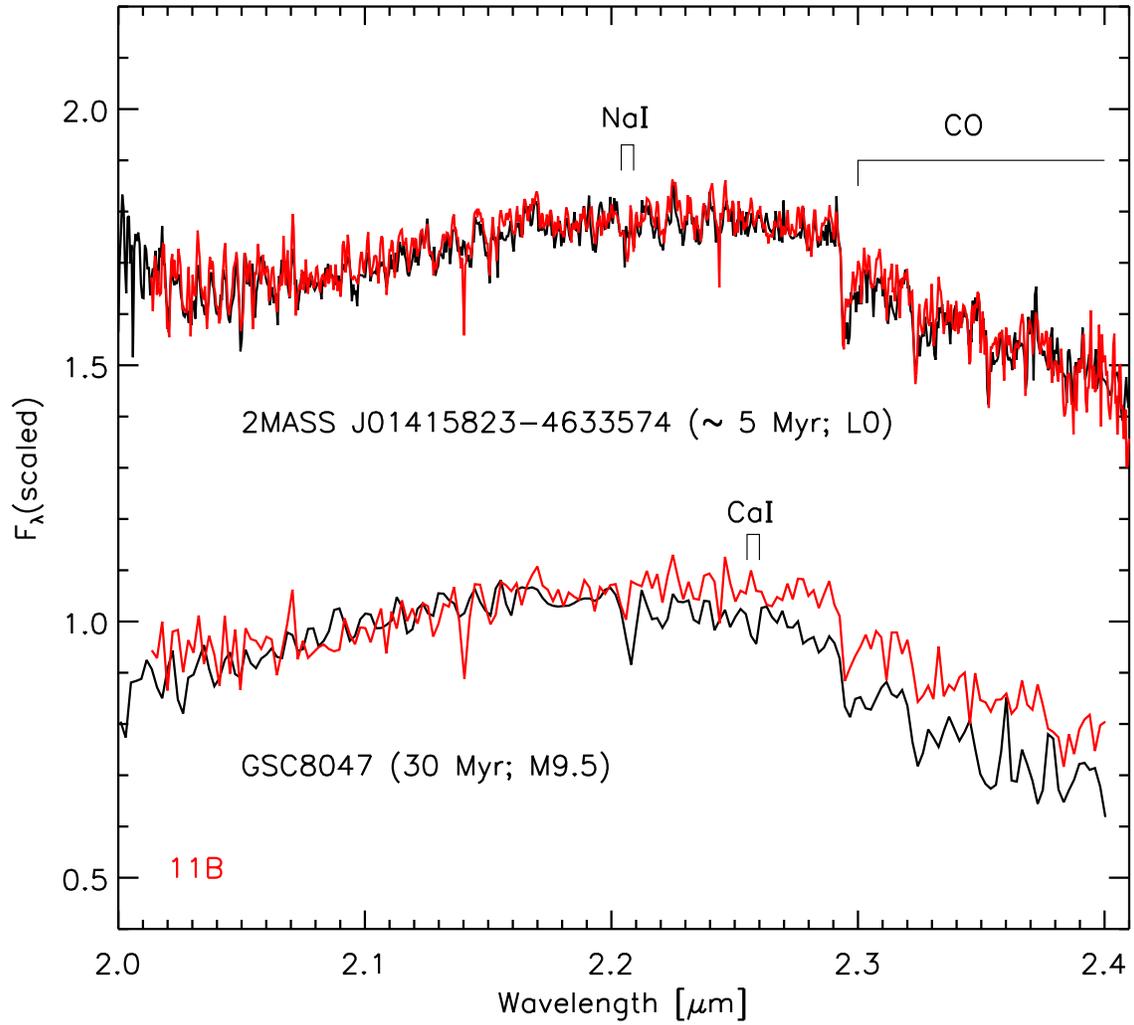}
\caption{
Similar to Figure 5, except for 11B. Note the even better fit to
 2M0141 for 11B. Hence we suggest that 11B is similar to a young
 M$9.5\pm0.5$, consistent with the optical spectral type of
 M9.5-L0 found by Jayawardhana \& Ivanov (2006b).  }
\label{fig6}
\end{figure}

\begin{figure}
 \includegraphics[angle=0,width=\columnwidth]{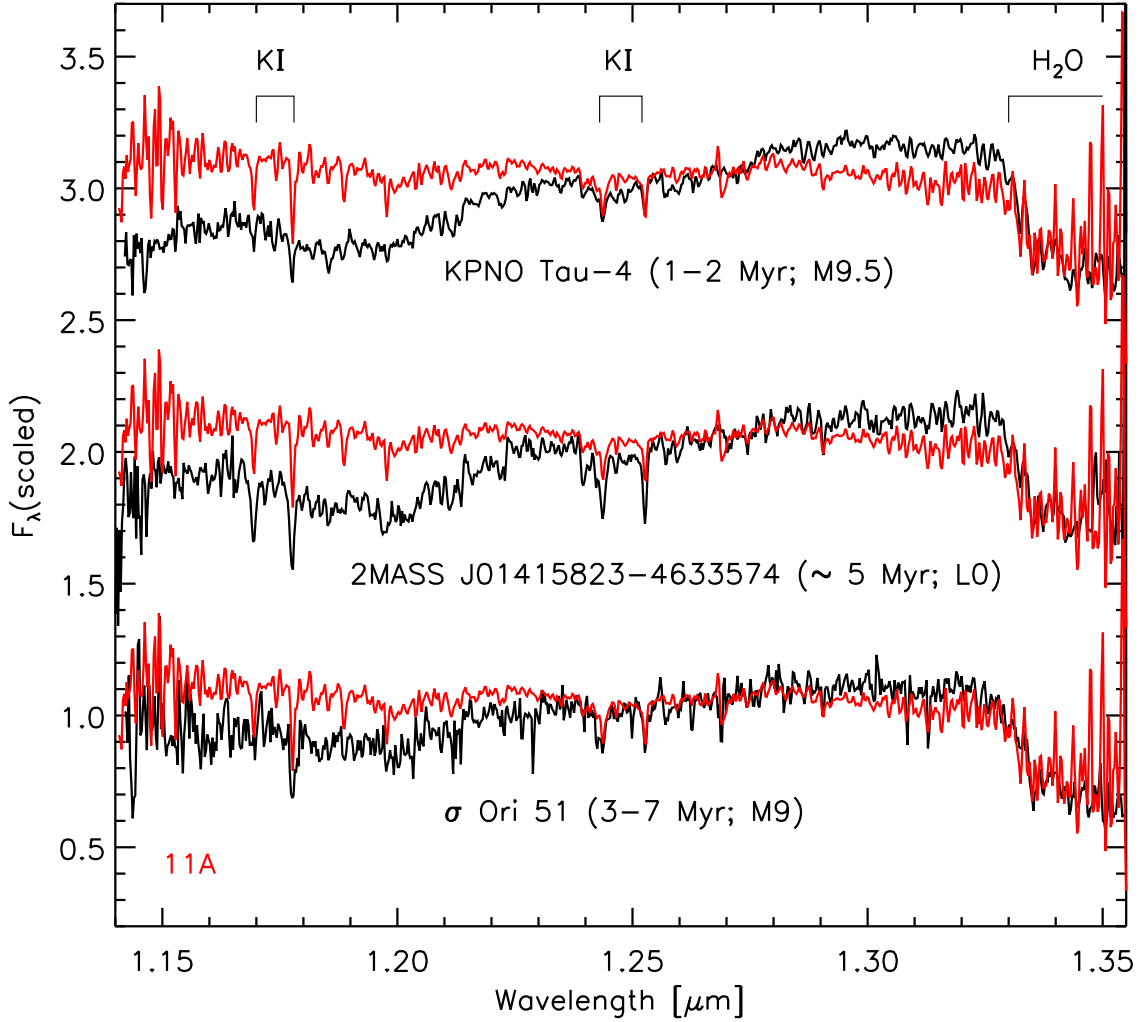}
\caption{
The J-band NIRSPEC spectrum of Oph 11A compared to a range of young
 brown dwarfs. Here we see the many gravity sensitive features in the
 J-band (McGovern et al. 2004) can be used to estimate the gravity
 (and hence age) of 11A. To minimize uncertainly we compare to other
 spectra obtained with the same instrumental set-up as we used. These
 J-band comparison spectra from the work of McGovern et al. (2004) and
 Kirkpatrick et al. (2006) show that 11A's KI doublets and J-band
 pseudo-continuum best fit a brown dwarf of age $5\pm2$ Myr such as
 $\sigma$ Ori 51 (McGovern et al. 2004). Moreover, the strength of
 11A's KI doublets (and poor fit to the pseudo-continuum) compared to
 the very young brown dwarf KPNO Tau-4 strongly suggests that the
 system age for Oph 11 is $>2$ Myr.  }
\label{fig7}
\end{figure}

\begin{figure}
 \includegraphics[angle=0,width=\columnwidth]{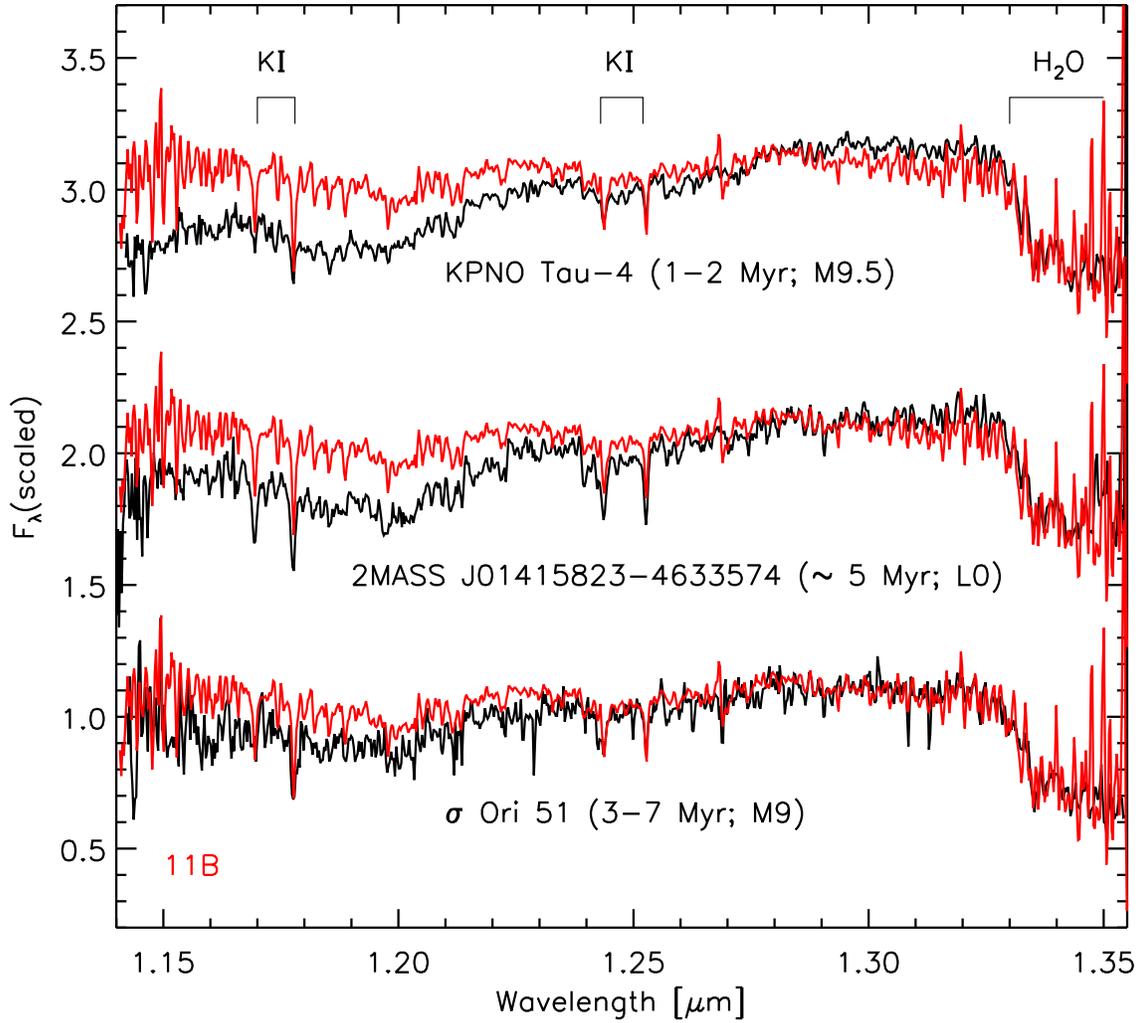}
\caption{
Similar to Figure 7 but with 11B. We see the surface gravity of the
 $5\pm2$ Myr $\sigma$ Ori 51 is the best fit. Moreover, the poor fit
 to KPNO Tau-4 strongly suggests that the system age for Oph 11 is
 $>2$ Myr.  }
\label{fig8}
\end{figure}

\begin{figure}
 \includegraphics[angle=0,width=\columnwidth]{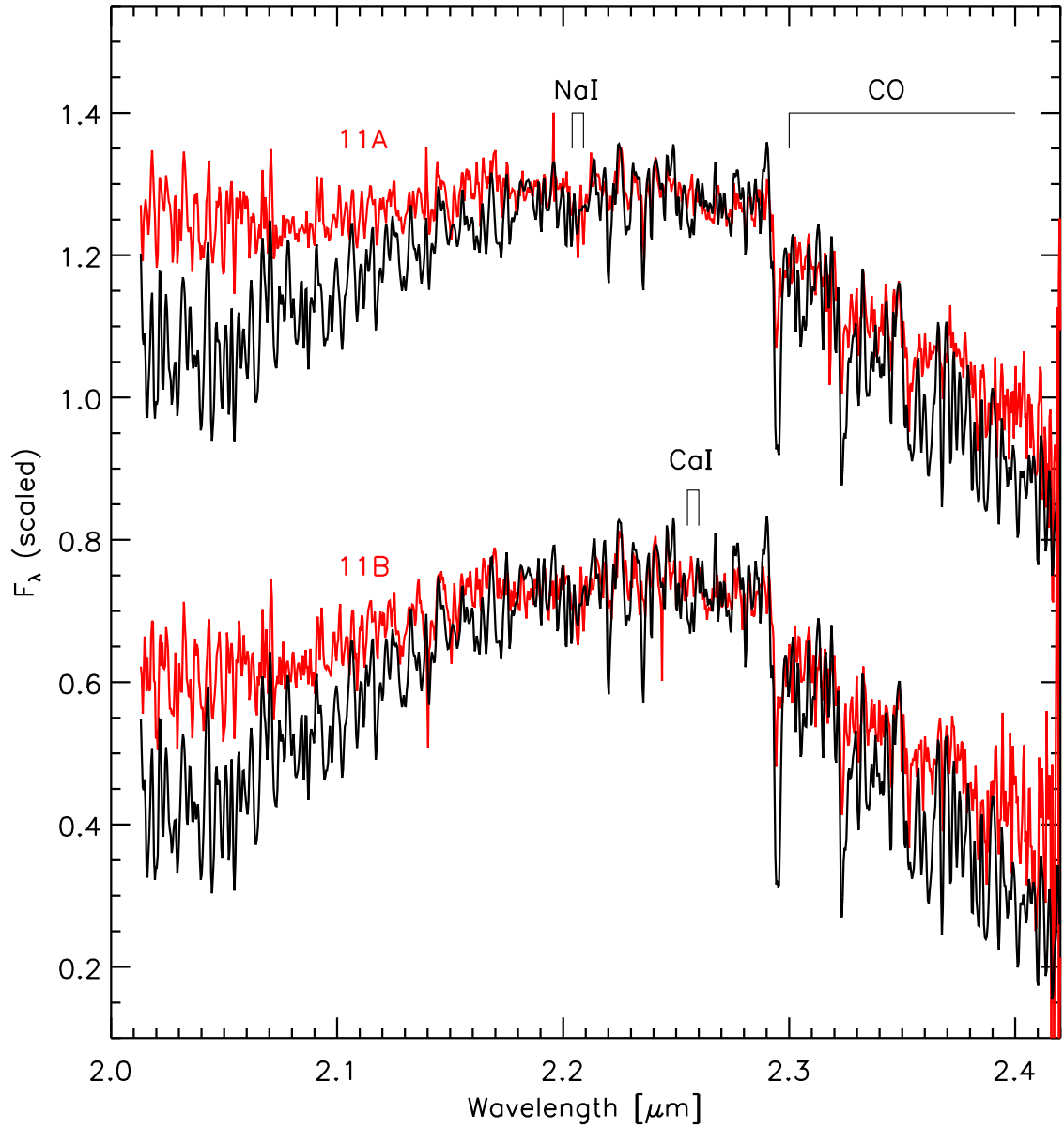}
\caption{
Comparison of our observed K-band spectra (red) to synthetic spectra
(black) computed using up-to-date {\tt PHOENIX} DUSTY atmosphere
models (Hauschildt et al. in prep).  See Section 3.3.1 for a discussion.}
\label{fig9}
\end{figure}

\begin{figure}
 \includegraphics[angle=0,width=\columnwidth]{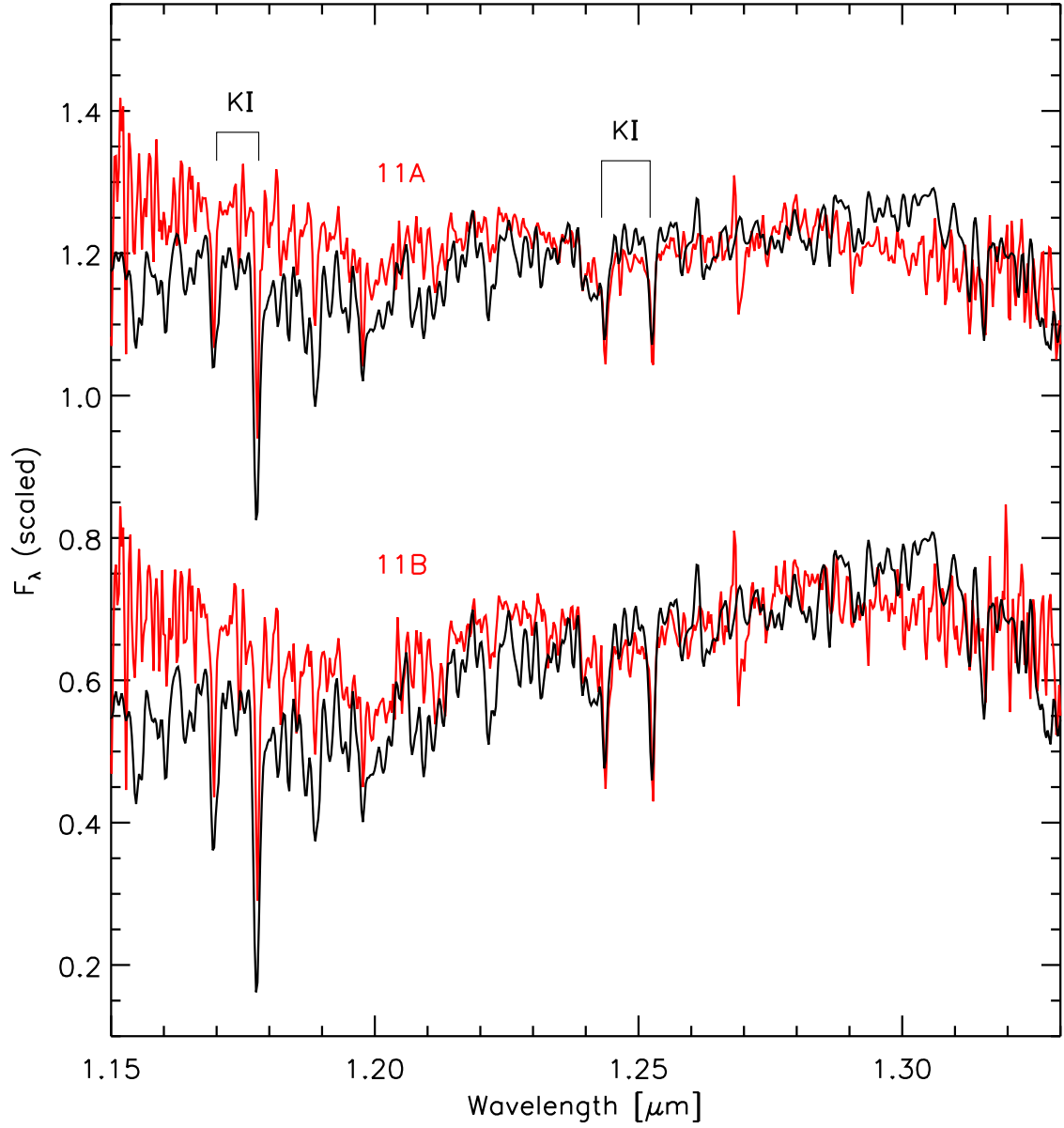}
\caption{
Same as Figure 9 except for the J-band instead of the K-band. See Section 
3.3.2 for a discussion.}
\label{fig10}
\end{figure}

\begin{figure}
 \includegraphics[angle=0,width=\columnwidth]{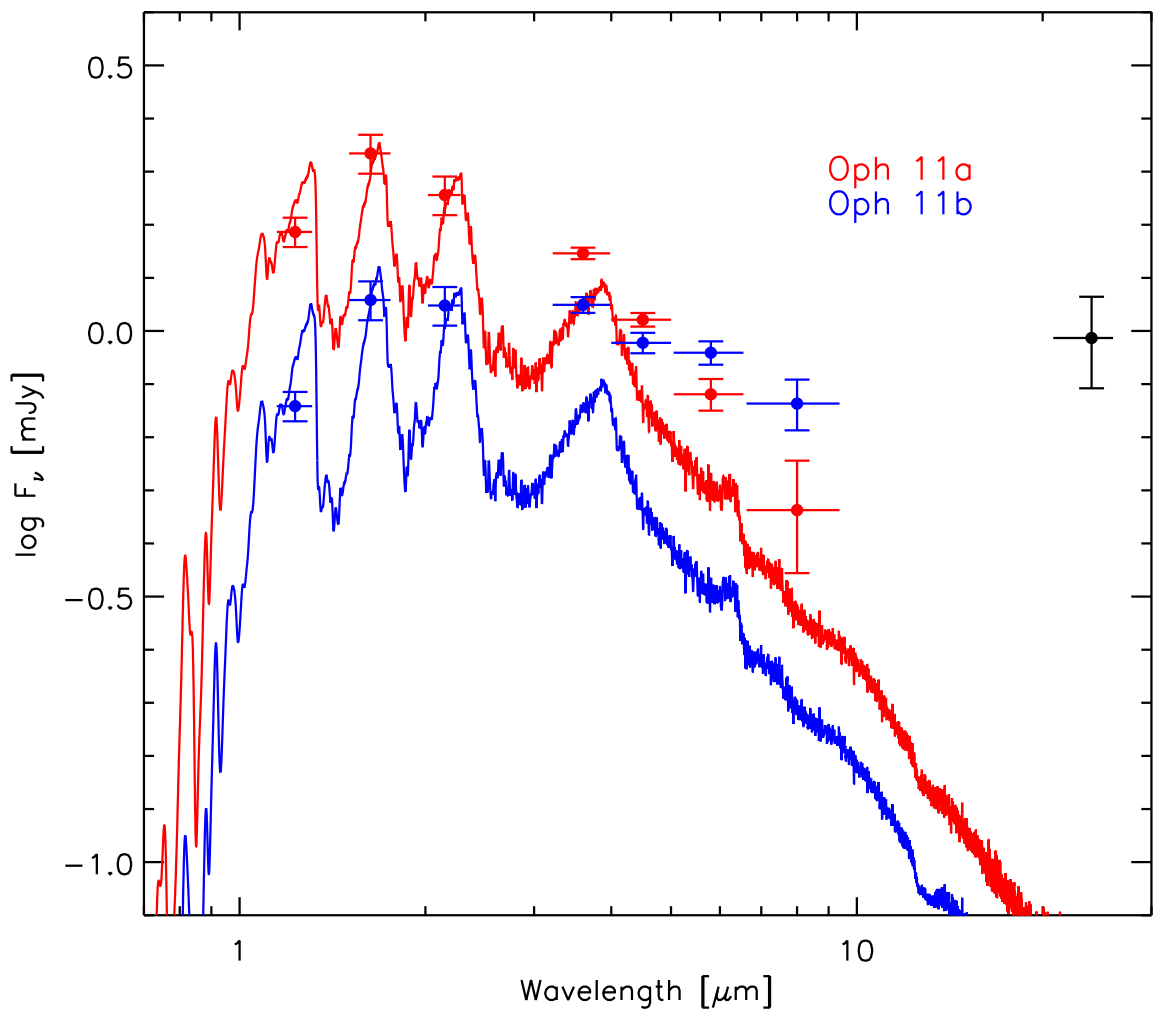}
\caption{
Synthetic 2MASS and IRAC fluxes for log(g)=3.95 and
$T_{eff}=2375$K and $T_{eff}=2175$K. These SEDs were computed using
up-to-date {\tt PHOENIX} DUSTY atmosphere models (Hauschildt et al. in
prep), assuming 125 pc as the distance to Oph \#11. These SEDs 
demonstrate the significant IR excess for 11A (red points) and 11B
(blue points) at $\lambda > 3
\mu$m. Component 11B appears to have a much stronger excess than 11A. The integrated 24$\mu$m datapoint for the system is plotted since we could not resolve the system at 24$\mu$m. Overall, the strong warm dust excess and H$\alpha$ emission suggests ages $<8$ Myr, consistent with our estimated $5\pm2$ Myr age. }
\label{fig11}
\end{figure}

\begin{figure}
 \includegraphics[angle=0,width=\columnwidth]{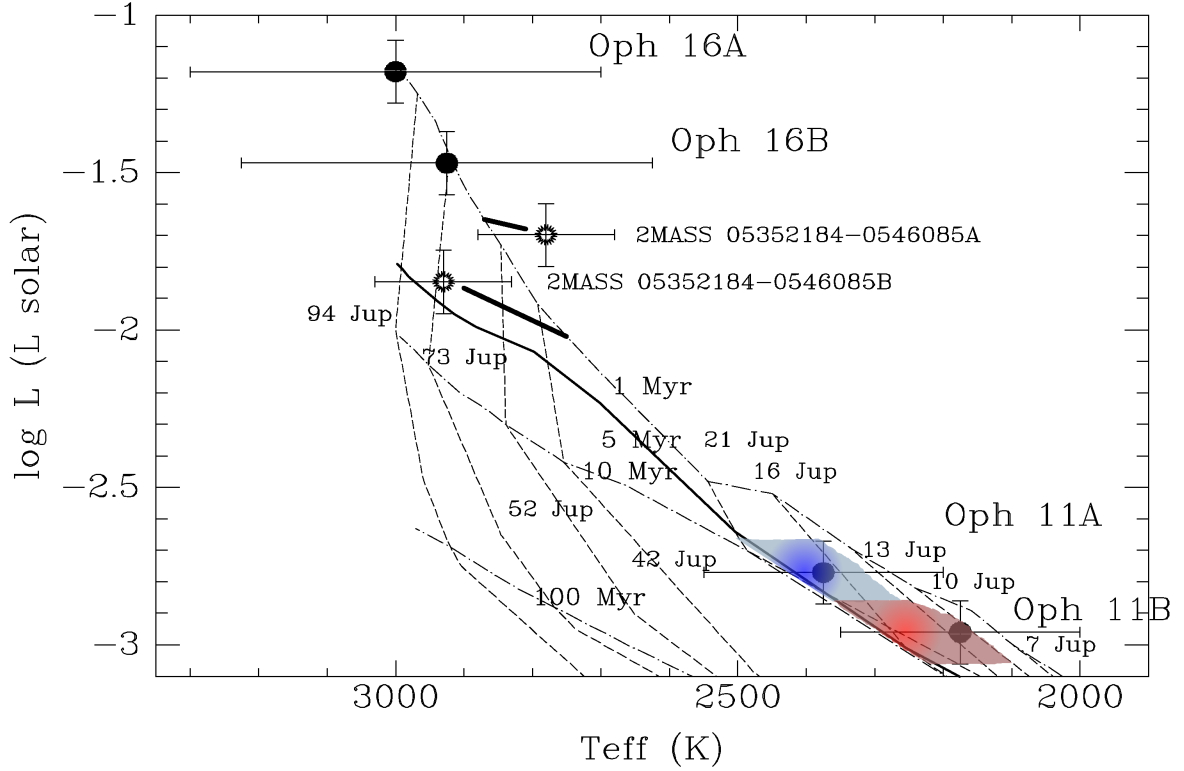}
\caption{
An HR diagram where the dashed ``vertical'' lines are iso-mass contours
for the DUSTY models (from left to right: 94, 73, 52, 42, 21, 16, 13, 10,
\& 7 M$_{Jup}$), while the four slightly more ``horizontal'', lines are the
DUSTY isochrones (top to bottom: 1 Myr (dash-dot) and 5 Myr (solid
line), 10Myr (dash-dot) and 100 Myr (dash-dot)).  The 2MASS 0535 data points (open circles) mark the
only known low-mass, very young, binary with a well determined
dynamical mass. The thick short diagonal lines representing the
displacement from the measured luminosity and temperature to the
values actually predicted by the DUSTY models. One can see that for
the secondary of the 2MASS 0535 eclipsing binary system (in Orion;
Stassun et al. 2006) the models have some error, but in general the
binary is close to the relevant 1 Myr isochrone. Hence we can have
some confidence in these models to estimate masses. From the
intersection of the 11A  (blue area) and 11B (red area) areas w.r.t. the estimated age of
$3-7$ Myr (from our NIR surface gravity measurements) we 
estimate masses of $13-21 M_{Jup}$ and $10-20 M_{Jup}$ for
11A and 11B. Therefore, the system appears to be a very low-mass brown
dwarf binary.  }
\label{fig12}
\end{figure}

\begin{figure}
 \includegraphics[angle=0,width=\columnwidth]{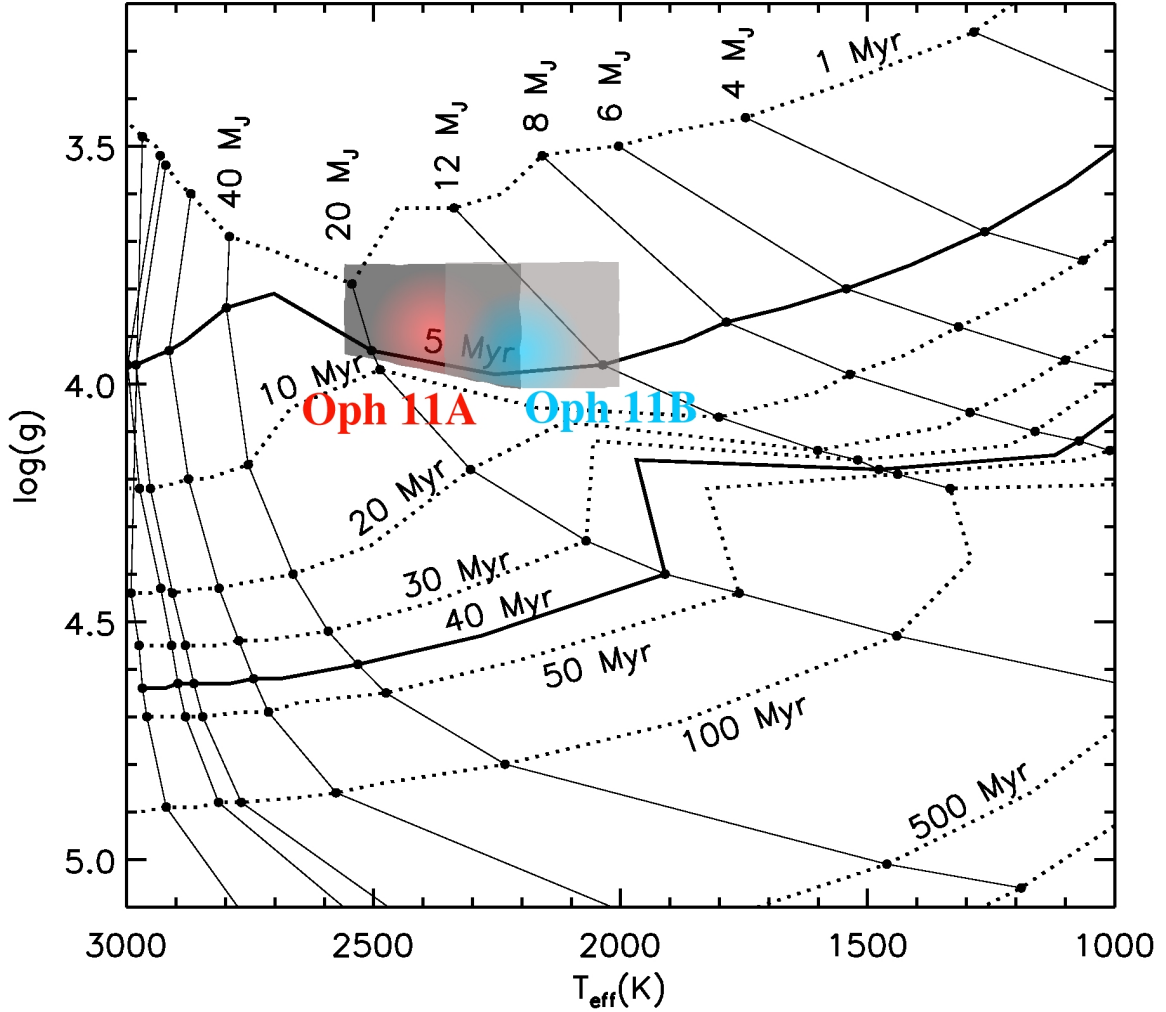}
\caption{
DUSTY tracks of Chabrier et al. (2002) in terms of
surface gravity (log(g) in cgs units) against temperature
($T_{eff}$). Hence, we can estimate masses independent of the possibly
variable NIR luminosity of Oph 11. From our synthetic spectral fits
(with the same models) we find log(g)$>3.75$. Moreover, the existence
of a strong IR excess and H$\alpha$ emission (Jayawardhana \& Ivanov
(2006b)) suggests that the age of Oph 11 is $<8$Myr. Therefore, we
``shade-in'' the remaining areas of model space that are
self-consistent for 11A and 11B. For 11A (red ``area'') we find
$17^{+4}_{-5} M_{Jup}$ is the allowed mass range. For 11B (blue
``area'') we find $14^{+3}_{-5} M_{Jup}$ is the allowed mass range.  }
\label{fig13}
\end{figure}

\begin{figure}
 \includegraphics[angle=0,width=\columnwidth]{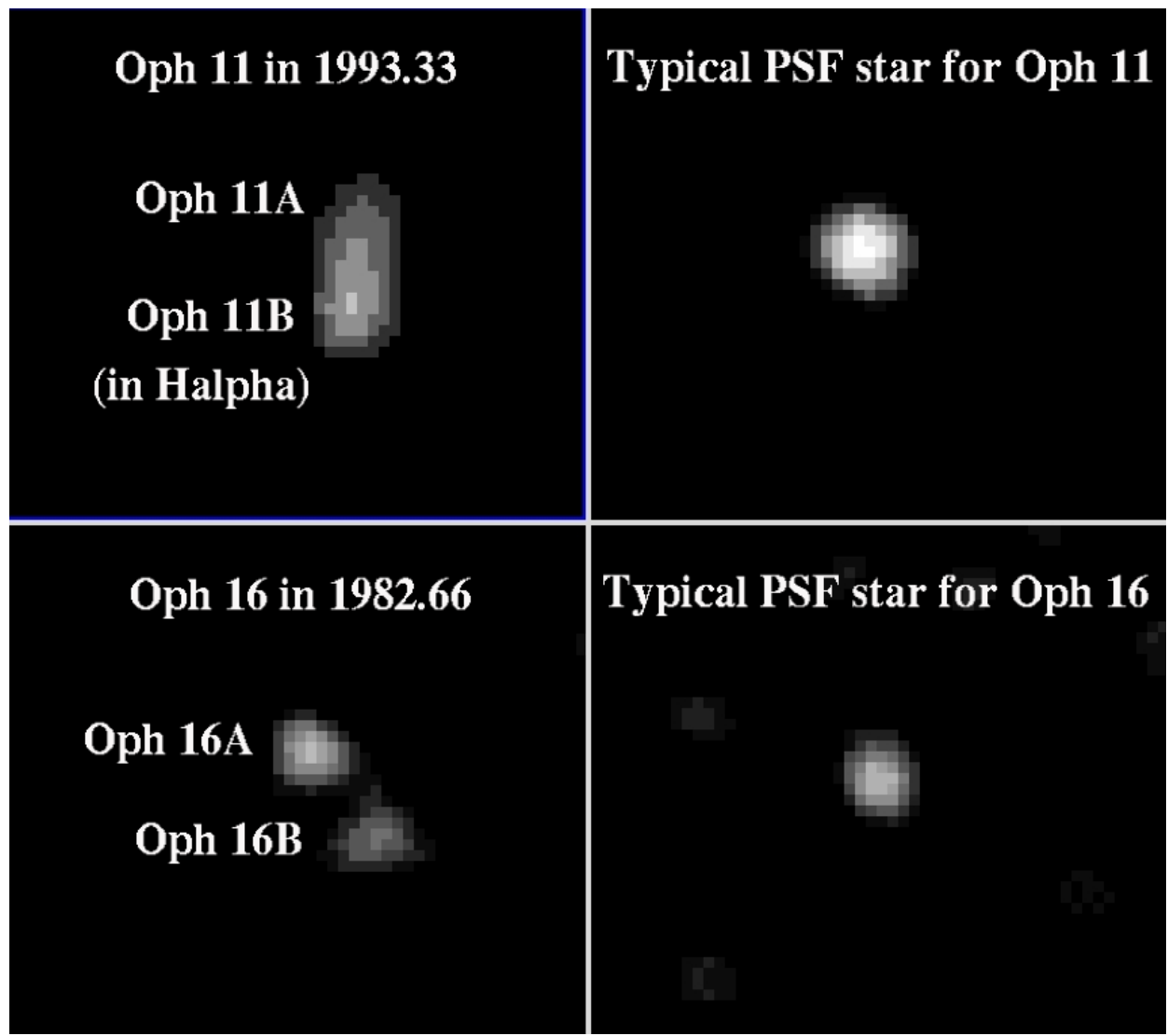}
\caption{
Upper left a 1993.33 epoch red POSS II image of Oph 11A
 and Oph 11B (11B is brighter than 11A in this filter due to its
 stronger H$\alpha$ emission). Note how nearby field PSF stars have
 circular PSFs very different from the Oph 11 system. This image shows
 the system with a separation of $2.05\pm0.20\arcsec$ and a
 $PA=174\pm6^{\circ}$ which is consistent with the current (2006.55)
 orientation and separation of the binary. Hence, we measure only
 $\sim3\pm5$ km/s of ``motion'' in the plane of the sky over the last 13.22
 years, consistent with the system being a bound, common
 proper motion, pair. This image has been unsharp masked, and
 magnified 5x5. The bottom panels show the IR POSS II image of the
 Oph 16 system in 1982.66. We find in 1982.66 the system had a
 separation of $1.87\pm0.2\arcsec$ and PA= $216.3\pm10^{\circ}$, this
 position is consistent with its current position 23.89 years
 later. This implies the Oph 16 system is likely also a bound common
 proper motion system.  North is up and east is left in all images.  }
\label{fig14}
\end{figure}

\begin{figure}
\includegraphics[angle=0,width=\columnwidth]{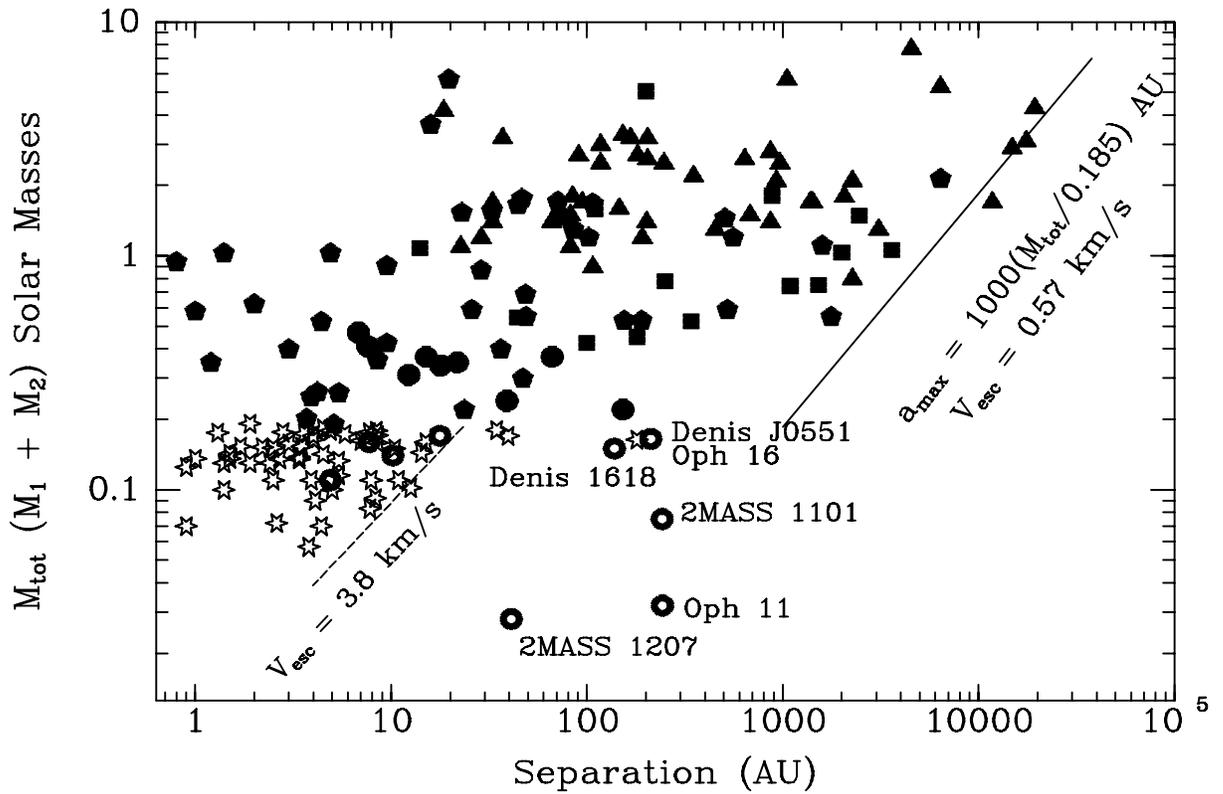}
\caption{
The Oph 11 system is compared to other known VLM and
brown dwarf binaries (old VLM systems are open stars, open circles are
young ($<10$ Myr) VLM systems, and stellar mass binaries are solid
symbols). Adopted from Close et al. (2003) and Burgasser et
al. (2006).}
\label{fig15}
\end{figure}

\begin{figure}
\includegraphics[angle=0,width=\columnwidth]{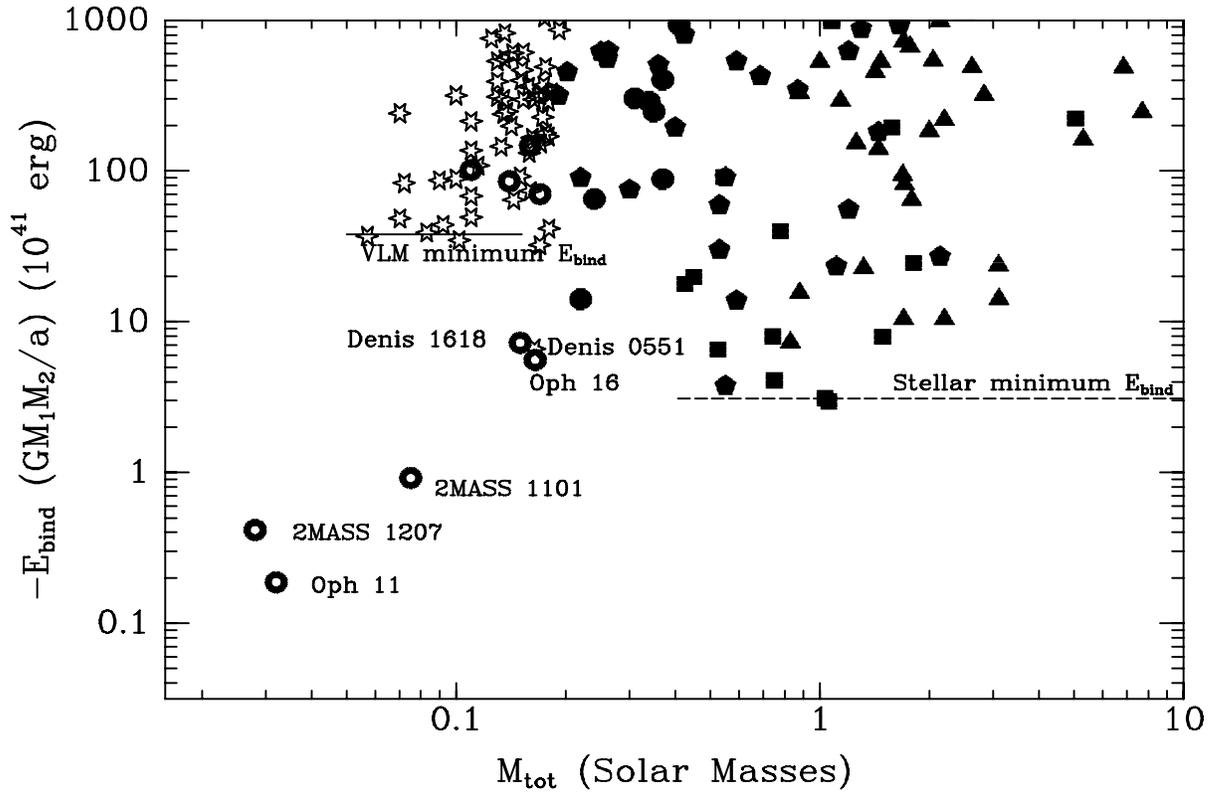}
\caption{
Illustrating the uniquely low binding energy of the Oph \#11 system
 compared to other known VLM systems. Note how our newly discovered Oph
 \#16 binary is also weakly bound. Adopted from Close et
 al. (2003) and Burgasser et al. (2006).  }
\label{fig16}
\end{figure}

\begin{figure}
\includegraphics[angle=0,width=\columnwidth]{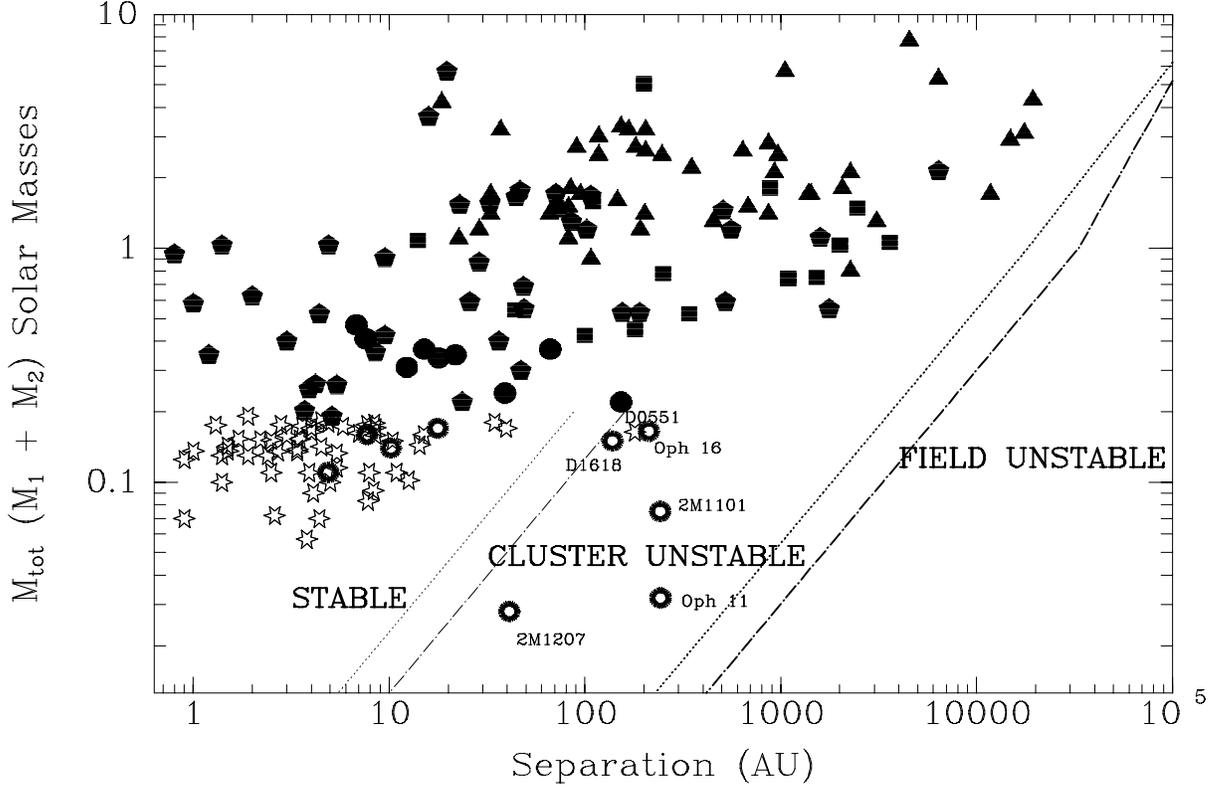}
\caption{ Same as Fig. 15, but with the instability zones
highlighted. The zones are determined by equations 1--4 which predict
the maximum bound separations plotted as dashed ``diagonal'' lines
from left to right as $sep^{diffusive*}_{cluster}$,
$sep^{catastrophic*}_{cluster}$, $sep^{diffusive*}_{field}$, and
$sep^{catastrophic*}_{field}$.  Note how all young ($<10$ Myr; open
circles), wide, VLM systems are in the ``Cluster Unstable''
region. Therefore, we expect that many of these ``cluster'' ($n_{*}\sim1000/pc^3$) binaries could be
evaporated before joining the field (open stars $>100$ AU). Hence,
wide VLM binaries in the field should be rare --as observed. We suggest that only wide VLM systems in very low ($n_*\la100/pc^3$) density groups will survive and join the field.  }
\label{fig17}
\end{figure}

\end{document}